\documentclass[a4paper,fleqn,usenatbib]{mnras} 

\pdfoutput=1
\usepackage[T1]{fontenc}
\usepackage{ae,aecompl}
\usepackage[export]{adjustbox}
\usepackage{graphicx}
\usepackage{amsmath}
\usepackage{amssymb}

\newcommand{\msun}{{\,\rm M_\odot}}
\newcommand{\vmax}{{\,$V_{\rm max}$~}}

\newcommand{\kms}{\,{\rm km}\,{\rm s}^{-1}}

\newcommand{\Mpc}{\,{\rm Mpc}}

\def\gsim{ \lower .75ex \hbox{$\sim$} \llap{\raise .27ex \hbox{$>$}} }
\def\lsim{ \lower .75ex \hbox{$\sim$} \llap{\raise .27ex \hbox{$<$}} }

 \title[The first galaxies in ETHOS]{ETHOS - an effective theory of structure formation: formation of the first haloes and their stars}
\author[M.~R.~Lovell et al.]{Mark R. Lovell\thanks{email: lovell@hi.is}$^{1,2}$, Jes\'us Zavala$^{1}$ and Mark Vogelsberger$^{3}$\thanks{Alfred P. Sloan Fellow}\\
$^{1}$Center for Astrophysics and Cosmology, Science Institute, University of Iceland, Dunhagi 5, 107 Reykjavik, Iceland \\
$^{2}$Institute for Computational Cosmology, Durham University, South Road, Durham DH1 3LE, UK\\
$^{3}$Department of Physics, Kavli Institute for Astrophysics and Space Research, Massachusetts Institute of Technology, Cambridge, MA 02139, USA}

\date{Accepted *** Received ***; in original
  form ***} 
\pubyear{2017}

\begin{document}

\label{firstpage}
\pagerange{\pageref{firstpage}--\pageref{lastpage}} 
  
\maketitle

\begin{abstract}
A cutoff in the linear matter power spectrum at dwarf galaxy scales has been shown to affect the abundance, formation mechanism and age of dwarf haloes and their galaxies at high and low redshift. We use hydrodynamical simulations of galaxy formation within the ETHOS framework in a benchmark model that has such a cutoff, and that has been shown to be an alternative to the 
cold dark matter (CDM) model that alleviates its dwarf-scale challenges. We show how galaxies in this model form differently to CDM on a halo-by-halo basis, at redshifts $z\ge6$. We show that ETHOS haloes at the half-mode mass scale form with 50~per~cent less mass than their CDM counterparts due to their later formation times, yet they retain more of their gas reservoir due to the different behaviour of gas and dark matter during the monolithic collapse of the first haloes
in models with a galactic-scale cutoff. As a result, galaxies in ETHOS haloes near the cutoff scale grow rapidly between $z=10-6$ and by $z=6$ end up having very similar stellar masses, higher gas fractions and higher star formation rates relative to their CDM counterparts. 
We highlight these differences by making predictions for how the number of galaxies with old stellar populations is suppressed in ETHOS for both $z=6$ galaxies and for gas-poor Local Group fossil galaxies. Interestingly, we find an age gradient in ETHOS between galaxies that form in high and low density environments.
\end{abstract}


\section{Introduction}
\label{intro}

The nature of the dark matter can be tested in a variety of different astrophysical environments, particularly those in which the masses of galaxies are close to the threshold for galaxy formation and thus close to a possible cutoff in the linear matter power spectrum as invoked by free-streaming models of warm dark matter \citep{Bode01,Boyarsky09a} and also by models where dark matter interacts with relativistic particles in the early Universe 
\citep{Boehm02,Buckley14,francis16,Vogelsberger16,Bose2018}. Such models are of interest because they have been shown to explain astrophysical phenomena as well as, and perhaps even better than, cold dark matter (CDM) models even in the presence of baryon physics processes, e.g. the densities of dwarf spheroidal galaxies \citep{Lovell12,Vogelsberger16,Lovell17a,Lovell17b}, the abundance of faint galaxies \citep{Schneider17}, and the ages of Local Group satellites \citep{Lovell17b}.

The focus of such tests to find signatures of new dark matter physics has typically been restricted to the abundance and properties of nearby galaxies in the Local Group, studies of which benefit from the ability to resolve stellar populations and measure kinematics but suffer from constraints of poor statistics and the stochasticity of halo assembly \citep{Polisensky2011,Lovell12,Boehm14,Lovell14,Polisensky14,horiuchi2016,Vogelsberger16,Bose17a,Lovell17a,Lovell17b,Schneider17,Bozek18}. 

At slightly higher redshift, current observational inferences of the $z=[0.2,0.5]$ isolated elliptical galaxies' subhalo mass functions provide only very weak constraints on models with a power spectrum cutoff \citep{Ritondale18,Vegetti18}. Analysis of the degree of substructure in the Lyman-$\alpha$ forest around $z=5-5.5$ have claimed the strongest constraints \citep{Viel13,Irsic17}, but the WDM particle physics models favoured by, for example, the dark matter decay interpretation of the 3.55~keV line observed in clusters and M31 \mbox{\citep{Boyarsky14a,Boyarsky14c,Bulbul14}} are in agreement with the Lyman-$\alpha$ constraints \citep{Garzilli15,Garzilli18}.

An alternative, and complementary, approach is to model and observe galaxies in the very high redshift Universe ($z>6$) where galaxies are only just starting to form, are low mass, and have a good opportunity to express the underlying dark matter model with fewer systematic uncertainties, e.g. satellite stripping and multiple stochastic star formation episodes, albeit at the cost of having to observe galaxies at much greater distances. Models with a primordial power spectrum cutoff have been shown to influence the population of galaxies at high redshift,
having an impact in
the collapse time \citep{Lovell12}, abundance \citep{Bode01,Polisensky2011,Lovell14,Kennedy14,horiuchi2016,Lovell16}, and UV luminosity of such galaxies \citep{Bose17a,Lovell18a,Menci18}. Models with cutoffs at progressively larger scales delay the collapse time, reducing the galaxy abundance and increasing the UV luminosity as the rapid, monolithic collapse of haloes leads to bright starbursts \citep{Bose16c,Lovell18a}. These studies have also shown that such models successfully reionize the universe within the constraints set by the Planck mission \citep{PlanckCP15} and high redshift galaxy counts \citep[e.g.][]{Bouwens15,Robertson15}. Thus, they remain compatible with current observations and at the same time leave a distinct imprint of new dark matter physics in the galaxy population, as opposed to the CDM model, which remains
featureless at all scales of interest in the physics of galaxies.

In this study we follow up on our previous paper, \citet{Lovell18a}, by breaking down the populations of simulated galaxies generated therein and examining them on an individual basis. In particular, we match haloes between our ETHOS simulation and its CDM counterpart. In the first instance this study will inform us about the mass scale at which such a one-to-one relationship between CDM and ETHOS haloes breaks down, in the sense that a CDM halo is or is not replicated by the ETHOS initial conditions. For those haloes that do have a one-to-one match, we will compare the halo masses and stellar properties such as star formation histories. 

This paper is organised as follows. In Section~\ref{sims} we review our simulations and methods, in Section~\ref{res} we present our results, and draw conclusions in Section~\ref{conc}.
 
\section{Simulations and methods} 
\label{sims}

 The simulations used in this paper are the same as performed for the previous study in this series \citep{Lovell18a}. We provide a summary here.
 
\begin{figure}
	\includegraphics[scale=0.33]{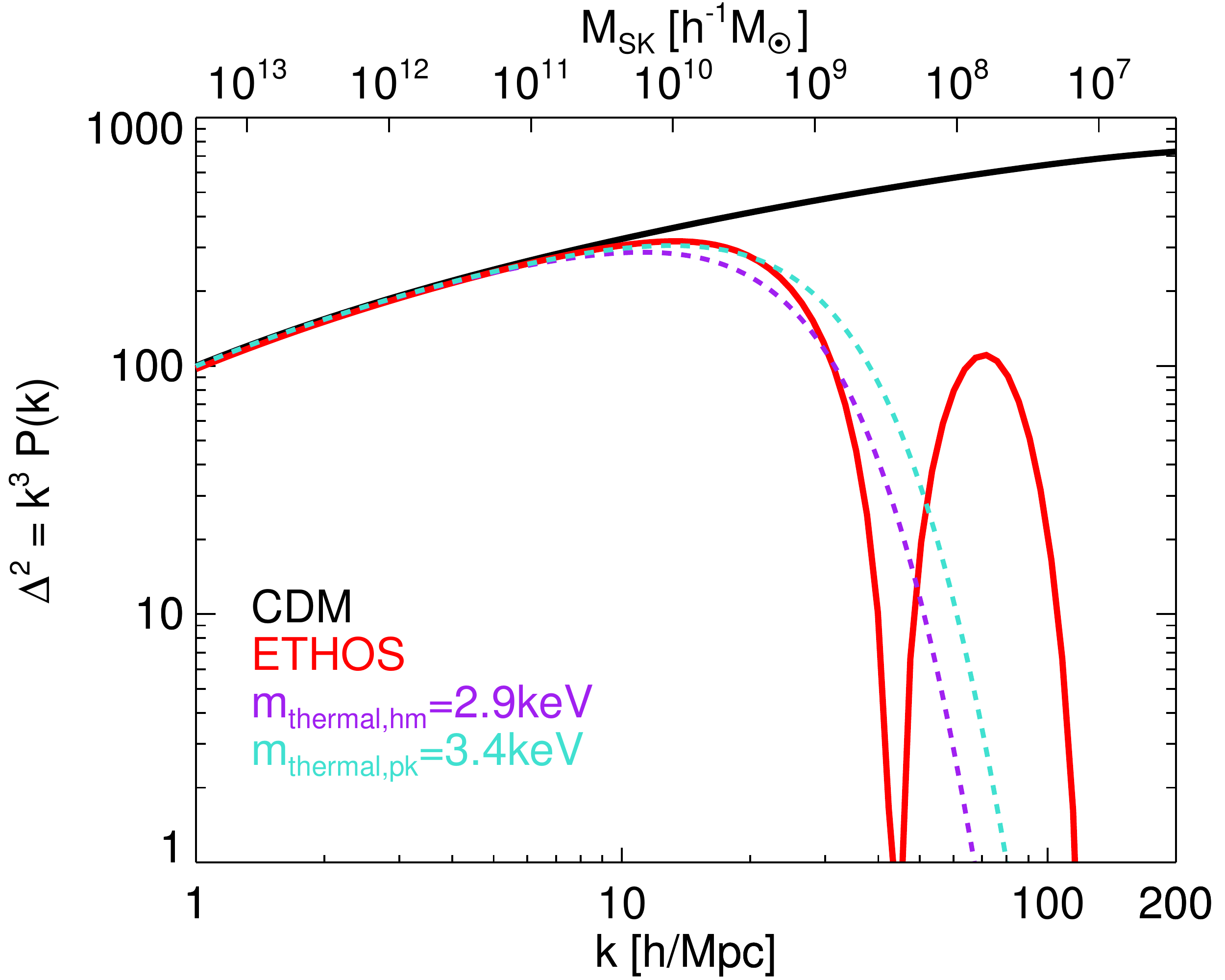}
    \caption{The dimensionless linear matter spectra for the ETHOS (red) and CDM (black) simulations used in this work. We also include two thermal relic models as dashed lines: that which has the same half-mode mass as ETHOS (2.9~keV, purple) and also the spectrum that has the same peak position as ETHOS (3.4~keV, turquoise).}
    \label{Pwspm}
\end{figure}

The simulations were performed using the hydrodynamical {\sc Arepo} code \citep{Springel10} and a leading model of galaxy formation \citep[][]{Vogelsberger13, Torrey14, Vogelsberger14,Vogelsberger14b,Genel14, Weinberger17, Pillepich17}. The code includes a module that implements elastic and isotropic self-interactions with velocity-dependent cross-sections \citep{Vogelsberger12,Vogelsberger16}. We note that this code has has been also employed for multiple other SIDM explorations~\citep[e.g.][]{Vogelsberger2013sidm, Creasey17, Brinckmann17}, and recently was also extended to include inelastic collisions~\citep[][]{Vogelsberger2018}. The benchmark model of interacting dark matter is the ETHOS-4 model (hereafter simply `ETHOS'), which is a specific model of the  ETHOS framework \citep{CyrRacine16}, that was calibrated in \citet{Vogelsberger16} to alleviate the CDM dwarf-scale challenges. It features dark matter self-interactions and a cutoff in the linear matter power spectrum due to dark matter$-$dark radiation interactions (see section~2 and table~1 of \citealt{Vogelsberger16} for details of the specific particle physics model and its parameters). The linear matter power spectrum of the ETHOS model is shown in Fig.~\ref{Pwspm}, alongside the CDM power spectrum and two thermal relic spectra, one of which has the same half-mode mass (defined below) as ETHOS (2.9~keV) and the other the same peak position in $k$ as ETHOS (3.4~keV). The cosmological parameters of our simulations are consistent with the {\it Planck} satellite estimates \citep{Planck14,Spergel15} and take the values: matter density $\Omega_{0}=0.302$,  dark energy density $\Omega_{\Lambda}=0.698$, baryon density $\Omega_{\rm b}=0.046$, Hubble parameter $H_0 = 100\,h\,\kms \Mpc^{-1} = 69.1\,\kms \Mpc^{-1}$,  power spectrum normalisation $\sigma_{8}=0.838$, and power spectrum slope index $n_\rmn{s}=0.967$. 

The cosmological boxes have a volume of $(36.2\Mpc)^3$, a dark matter particle mass of $1.76\times10^{6}\msun$ and an average gas cell mass of $2.69 \times 10^5\msun$. The softening lengths for the dark matter and the gas cells are 724~pc (comoving) and 181~pc (comoving minimum, adaptive) respectively. The simulations were evolved from $z=127$ to $z=6$. Haloes are identified using a Friends-of-Friends (FoF) algorithm, and gravitationally bound subhaloes are determined using the {\sc subfind} code \citep{Springel01a}. The process is repeated with the same initial conditions phases, volume and galaxy formation model to produce a CDM counterpart. 

In order to identify pairs of matching haloes between the CDM and ETHOS versions we employ the Lagrangian region matching method introduced in \citet{Lovell14} and augmented in \citet{Lovell18b}. To summarise, we measure the degree to which prospective pairs of haloes in the two simulations -- as defined by the particles assigned to them by {\sc subfind} -- share the same Lagrangian regions in the initial conditions. The quality of the match is determined by the parameter $R$, which is defined as the ratio of the cross-potential energy of the two matter distributions squared and the product of the two individual potential energies (see fig.~6 of \citealp{Lovell14}): for a perfect match $R=1$, and for progressively worse matches $R\to 0$. The best possible match for each halo in the ETHOS simulation is the halo in the CDM simulation that returns the highest value of $R$. We perform the calculation in both directions, i.e., ETHOS matches for CDM haloes as well as CDM matches for ETHOS haloes. If a CDM halo's best match in ETHOS is also the best match determined for that ETHOS halo, then the match is said to be {\it bijective}. We use bijective pairs of haloes in most of our CDM--ETHOS comparisons in this paper.

The vast majority of bijectively matched haloes with a halo virial mass $M_\rmn{200,CDM}$\footnote{Defined as the mass within the radius that encloses an overdensity of 200 times the critical density.}>$10^{9}\msun$ achieve a value of $R>0.95$, and therefore constitute high quality matches where the same halo has collapsed from the same initial conditions volume patch in the two simulations as discussed below in Fig.~\ref{MatchRate}. Below $10^{9}\msun$ the distribution in $R$ shifts progressively to lower $R$, suggesting that in this regime there are far more incidences where stochastic formation histories are becoming relevant. We conclude that relaxed CDM haloes with masses $>10^9\msun$ are reliably matched, and therefore reproduced, in the ETHOS simulation; we analyse the properties of lower mass haloes in the following Section. 

Subhaloes that form from spurious fragmentation of filaments are identified using the initial conditions sphericity method developed in \citet{Lovell14}. We comment in the text where these haloes may play a role in our results.

 \section{Results}
 \label{res}

 The goal of this study is to understand how the matter power spectrum cutoff affects the assembly of the first galaxies and their host haloes, and subsequently to make predictions for how these changes are expressed in the star formation histories of high redshift galaxies ($z>6$) and also present day fossil galaxies, i.e., galaxies that formed most of their stellar mass before the end of reionization at $z=6$. In this Section we therefore begin by showing how haloes are reproduced in the ETHOS simulation compared to CDM, then compare the halo and galaxy masses, and finally present results for the star formation histories.
 
 \subsection{Halo formation near the cutoff scale}

We begin our analysis with an examination of the formation of haloes in CDM and ETHOS, and show to what degree CDM haloes are reproduced with similar properties in ETHOS. We will focus on galaxies at two distinct epochs: $z=6$, which is at the end of reionisation and also the last snapshot output time for which we have simulation data, and secondly $z=10$ as a compromise between, on the one hand, the set of redshifts where CDM and ETHOS differed most in \citet{Lovell18a}, and on the other hand, the range of observational accessibility both at the present time and in the near future.

\subsubsection{Halo matching}

  Our haloes have been matched between CDM and ETHOS as described in Section~\ref{sims}. Here we consider the fraction of haloes that have bijective matches in the two simulations. 

It is inevitable that some of the haloes in our sample will not have a bijective match, and this can occur for several reasons. Those few haloes at the high mass end that do not have a match are typically undergoing a major merger, in which case the subhalo finder's definition of which particles belong to which of the two merging haloes can have a profound effect on the assignment of the Lagrangian region. At lower masses on the other hand, there are effects caused by the physical suppression in the abundance of ETHOS structures due to the primordial cutoff in the power spectrum, and the numerical artefacts caused by discreteness effects, in particular the confusion of structures caused by spurious haloes \citep{Wang07}.

In order to find out to what degree these issues affect halo-to-halo matches, we compute the ratio of two subsets of our haloes to the total number of haloes for which matches were attempted\footnote{Approximately 1~per~cent of CDM haloes with $M_{200}>10^{9}\msun$ did not achieve any potential matches due to mergers changing the value of the mass, and are therefore excluded from this analysis.}. The first subset is for high quality matches, defined as having $R>0.9$, and a second of likely matches, with $R>0.5$; the latter is defined from previous studies \citep{Lovell18b} which have shown that the `second best' match to a halo, i.e. the candidate match halo that has the second highest value of $R$ and is therefore the first `false' match, returns $R\le0.4$. 

\begin{figure}
 	\includegraphics[scale=0.35]{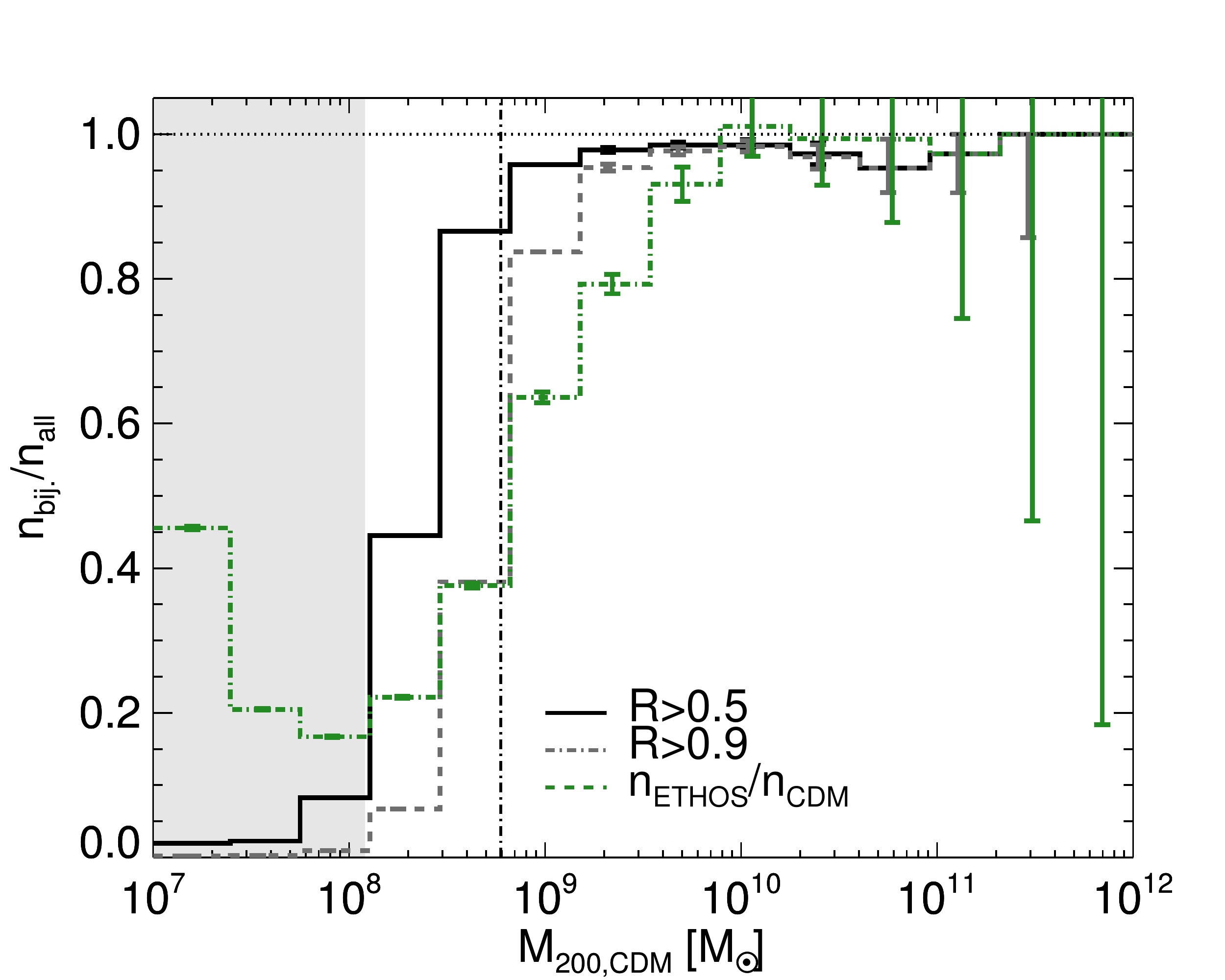}\\
 	\includegraphics[scale=0.35]{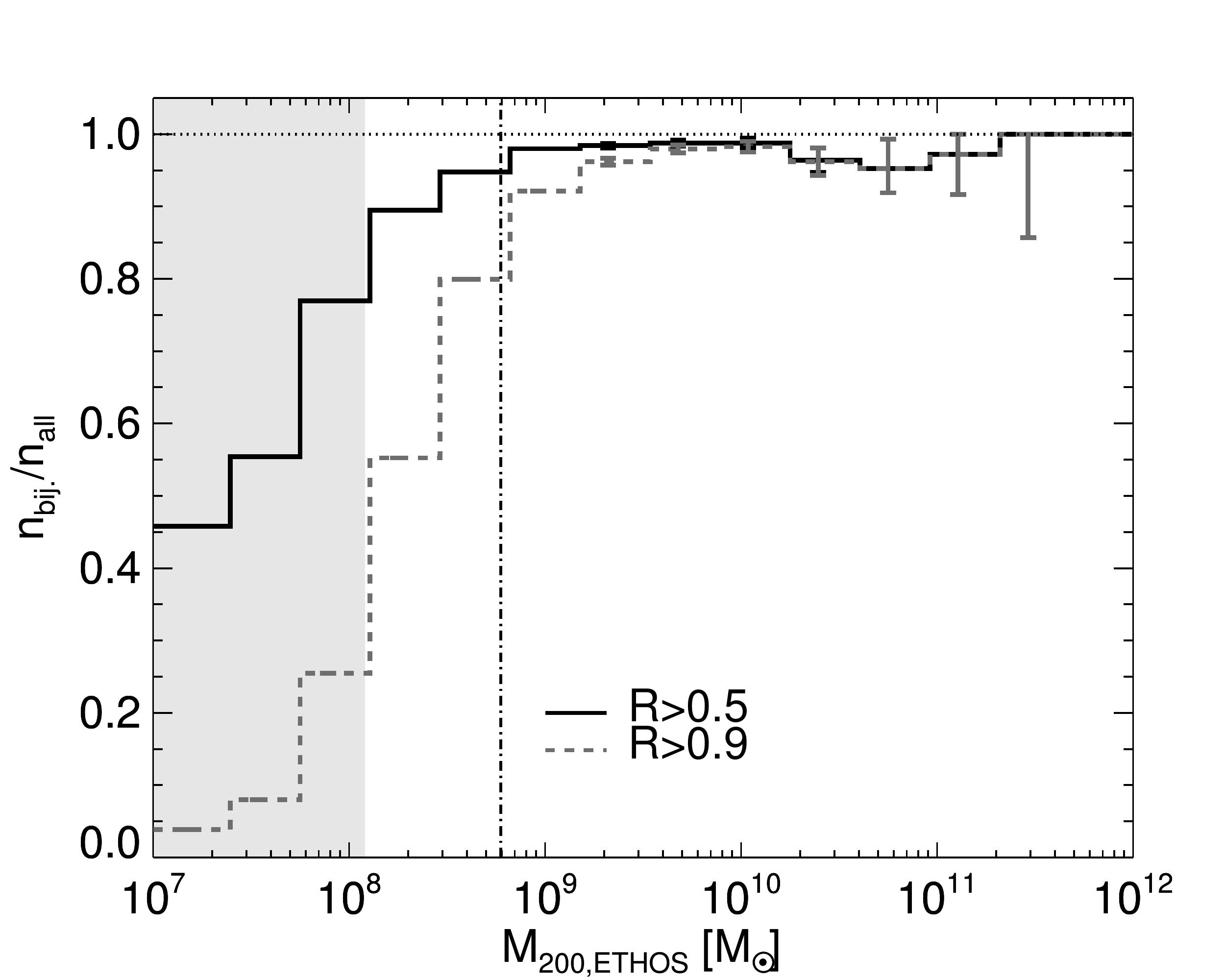}
    \caption{Fraction of haloes with {\it bijective} matching pairs (see last paragraphs of Section~\ref{sims}) as a function of $M_{200}$. The top panel shows the ratio of CDM bijectively matched haloes to all CDM haloes, and the bottom panel the same quantity for ETHOS. The solid black curves show the ratio when including haloes of $R>0.5$, and grey dashed curves where $R>0.9$, where $R$ is the value of the matching quality parameter described in Section~\ref{sims}. The dot-dashed green curve in the top panel shows the ratio of the total number of ETHOS haloes to CDM haloes in each bin, independent of matches. The error bars on the bijective match ratios are binomial ($2\sigma$) and the errors on the mass function ratios are Poissonian ($1\sigma$). The half-mode mass for ETHOS is shown as a vertical dot-dashed line. The grey region delineates where spurious bhaloes contribute to the mass function according to the \citet{Wang07} criterion.}
    \label{MatchRate}
 \end{figure}

We present our results in Fig.~\ref{MatchRate}, first for the rate at which CDM haloes have a bijective match in ETHOS (top panel) and second for the proportion of ETHOS haloes that have a bijective match in the CDM simulation (bottom panel). We also include the half-mode mass, $M_\rmn{hm}$, which is defined as the mass enclosed within the comoving scale at which the CDM-ETHOS transfer function is suppressed by 50~per~cent:

\begin{equation}
	M_\rmn{hm} = \frac{4\pi}{3}\bar\rho\left(\frac{\pi}{k_\rmn{hm}}\right)^{3}
    \label{eqn:Mhm}
\end{equation}

\noindent where $\bar\rho$ is the mean background density and $k_\rmn{hm}$ is the wavenumber at which the ETHOS linear matter power spectrum is suppressed by $0.5^2$ relative to the CDM power spectrum.  
 
 In all cases, the ratio of matched-to-total haloes is close to unity for $M_{200}>3\times10^{9}\msun$, including at the dip around $5\times10^{11}\msun$ where the presence of young, merging haloes is most likely to cause matching issues. For $M_{200}<10^{9}\msun$ both the $R>0.9$ and $R>0.5$ haloes start to peel off from a ratio of 1, the former much more sharply than the latter. The first bin at which there is a clear separation between the two samples is at $2\times10^9\msun$, where the value of the ratio for the high-quality sample is 15~per~cent lower than the likely match sample. This is therefore the regime at which stochasticity starts to affect the halo formation histories significantly. There is also a significant difference between the high quality matches for CDM and ETHOS at this mass scale: 80~per~cent of CDM haloes obtain a high quality match compared to 90~per~cent of ETHOS haloes, indicating that it is at this mass (around 50~per~cent higher than the half-mode mass) that the power spectrum cutoff is suppressing the halo abundance. At still lower masses the dropoff in the high quality match proportion is shallower in ETHOS. 

 Interestingly, this dropoff is much sharper, and also occurs at a lower halo mass, than would be inferred from comparing the two halo mass functions  as shown in Fig.~3 of \citet{Lovell18a}. We show this explicitly by plotting the ratio of the ETHOS to CDM mass functions in the top panel of Fig.~\ref{MatchRate}. This mass function ratio peels away from unity at a three times higher mass than the bijective match ratio. Finally, we note that at $10^{8}\msun$ the ratio of mass functions is higher than the matched proportion: we attribute this reversal to a combination of CDM haloes that have no match to ETHOS counterparts and ETHOS haloes that instead have matches to more massive CDM haloes. 

 \subsubsection{Halo mass}
 
\begin{figure}
 	\includegraphics[scale=0.44]{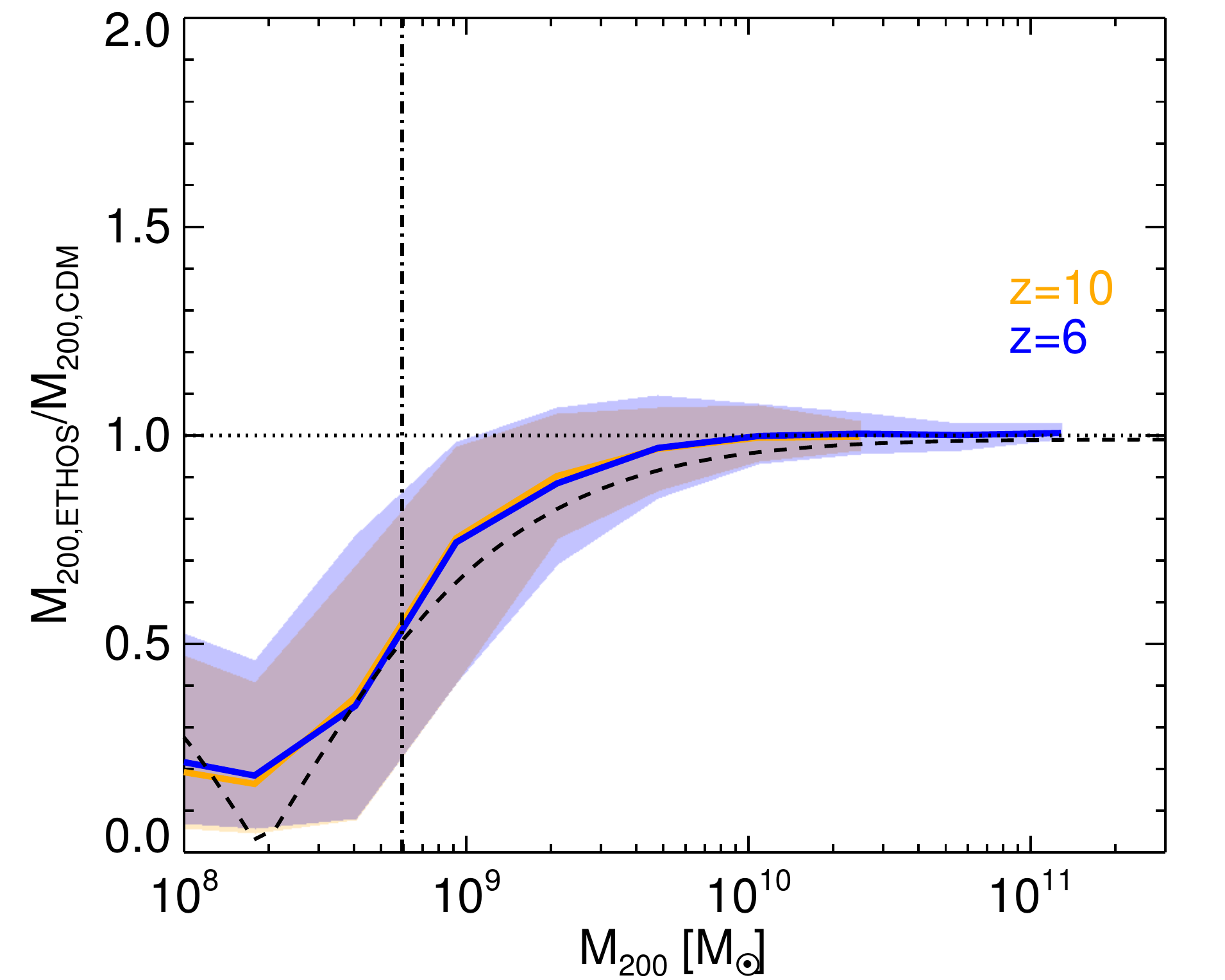}
    \includegraphics[scale=0.44]{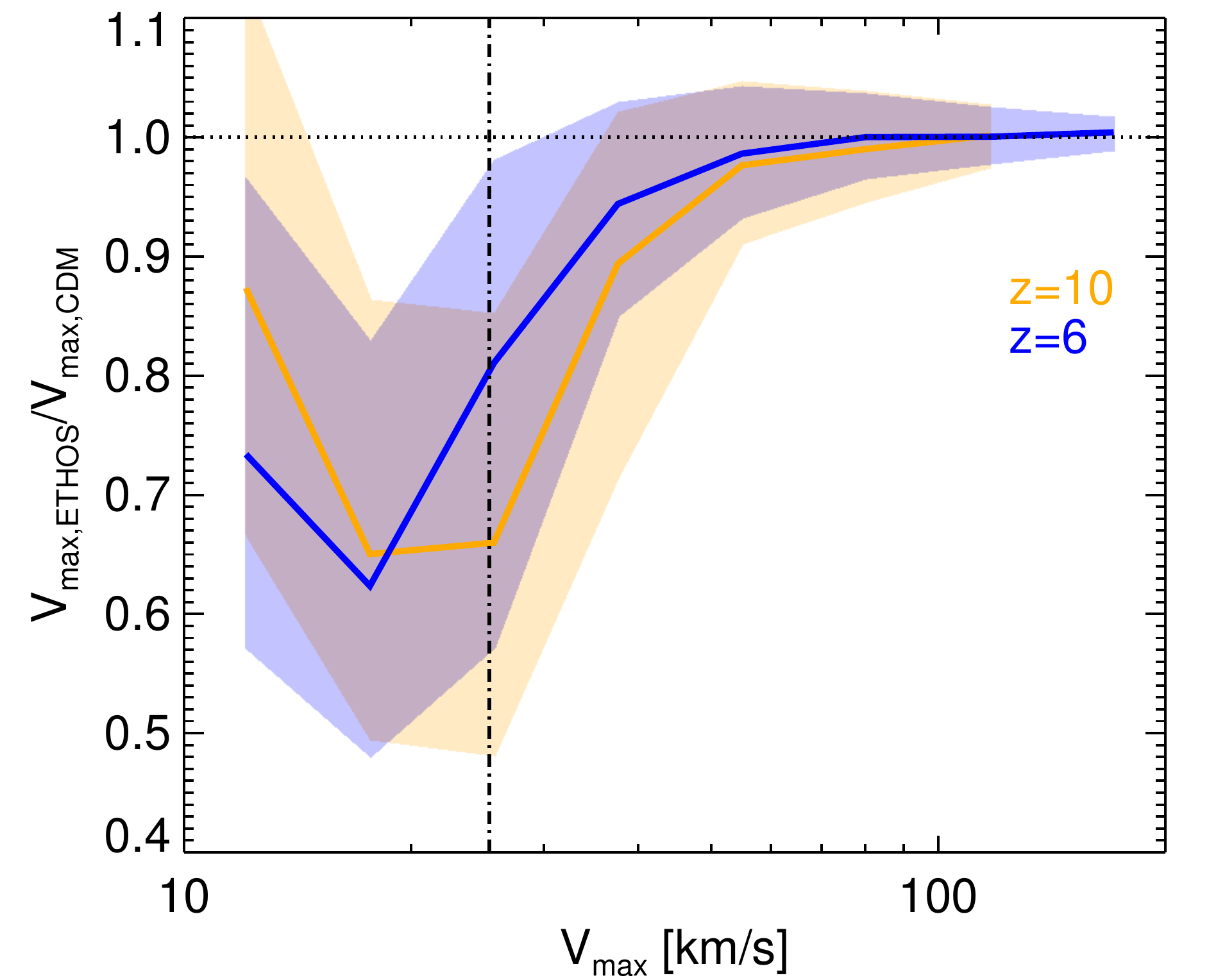}
    \caption{The ratio between halo virial masses $M_{200}$ (top panel) and the maximum circular velocities \vmax (bottom panel) of our bijective halo matches between the ETHOS and CDM models. The $z=6$ simulation data are shown in blue while $z=10$ is shown in orange. The solid lines show the median relations and shaded regions show the $1\sigma$ scatter of the distribution. The vertical dot-dashed line in the top panel shows the ETHOS half-mode mass, $M_\rmn{hm}$, and in the bottom panel it shows the half-mode $V_\rmn{max}$, defined as the median value of \vmax at $M_{200}=M_\rmn{hm}$ in CDM (at $z=6$). The dashed curve in the top panel shows the transfer function for the ETHOS power spectrum where the wavenumber $k$ has been mapped into $M_{200}$ by generalising Eq.~\ref{eqn:Mhm} for $k$ other than $k_\rmn{hm}$.}
    
    \label{M200Match}
 \end{figure}

We explore the source of the discrepancy between the mass functions and the matched fraction by means of our set of matched haloes ($R>0.5$ unless stated otherwise). For each pair we compute the ratio of the ETHOS halo $M_{200}$ to its CDM counterpart and then calculate the distribution of this halo mass ratio as a function of the CDM counterpart $M_{200}$. We plot the median and scatter of our results for redshifts $z=6$ and $z=10$ in Fig.~\ref{M200Match}. 
 
At $z=6$, the paired haloes have the same mass for $M_{200}>10^{10}\msun$. At lower masses the ETHOS halo masses are suppressed relative to their CDM counterparts, down to a minimum of 20~per~cent of the CDM counterpart mass at $10^8\msun$. An almost identical behaviour is observed at $z=10$. We therefore conclude that the discrepancy between the dark grey and green lines in Fig.~\ref{MatchRate}, i.e. the difference between the suppression of the halo mass function in ETHOS expected from object abundance alone to that which we measure, is due to the suppression of $M_{200}$ itself in ETHOS, even for haloes that exist in both simulations. 

We notice that the suppression of 50~per~cent in halo mass in the top panel of Fig.~\ref{M200Match} occurs almost exactly at the half-mode mas, $M_\rmn{hm}$, where the CDM-to-ETHOS transfer function is likewise suppressed by 50~per~cent. We therefore also  plot in the top panel of Fig.~\ref{M200Match} this transfer function as a function of $M_{200}$, where we relate the linear matter power spectrum / transfer function wavenumber $k$ to the halo mass via the same equation as used for $k_\rmn{hm}$ to $M_\rmn{hm}$ in Eq.~\ref{eqn:Mhm}. We find that the (linear) expectation from the primordial transfer function on the mass ratio between ETHOS and CDM is suppressed by up to 25~per~cent relative to the actual measurement in our simulations at $M_{200}>M_\rmn{hm}$.
 We will follow up in future work whether this approximation is appropriate for other cutoff shapes and redshifts.  

A similar result has been found in the past 
for the peak of the circular velocity curve, $V_\rmn{max}$ (which is defined less arbitrarily than the virial mass) in \citet{Bozek16} for $z=0$ WDM Local Group galaxies in DMO simulations (see their figure~9).  
We thus repeat our $M_{200}$ analysis for $V_\rmn{max}$ and present our results in the bottom panel of Fig.~\ref{M200Match}.
The results for $V_\rmn{max}$ closely mirror those of $M_{200}$, with large $V_{\rm max}$ haloes at both redshifts having very similar $V_\rmn{max}$ values in CDM and ETHOS, and a ratio that tapers down to a median suppression at $V_\rmn{max}\rmn{(CDM)}=20\kms$ of 20~per~cent (30~per~cent) for $z=6$ ($z=10$) haloes. This is slightly less severe than the 40~per~cent measured by \citet{Bozek16}, and may be due to the fact that we are looking at higher redshifts ($z=6-10$ as opposed to $z=0$) and also might reflect differences in the linear power spectrum cutoff between the ETHOS model we analysed and the family of WDM models analysed in \citet{Bozek16} which are approximated by a 2.0~keV thermal relic.
  
  The source of this suppression in mass ($V_{\rm max}$) in ETHOS relative to CDM is linked to the size of the overdensity, $\delta$, in which the haloes is centred, and subsequently to the concentration of the halo. A discussion of this topic is included in Appendix~\ref{sec:conc}.

 \subsection{The first galaxies in ETHOS}
 \subsubsection{Galaxy properties}
 
 We turn to the properties of the baryonic component of each galaxy, which we attempt to understand in the context of the change in halo mass discussed above.
 
\begin{figure}
 	\includegraphics[scale=0.45]{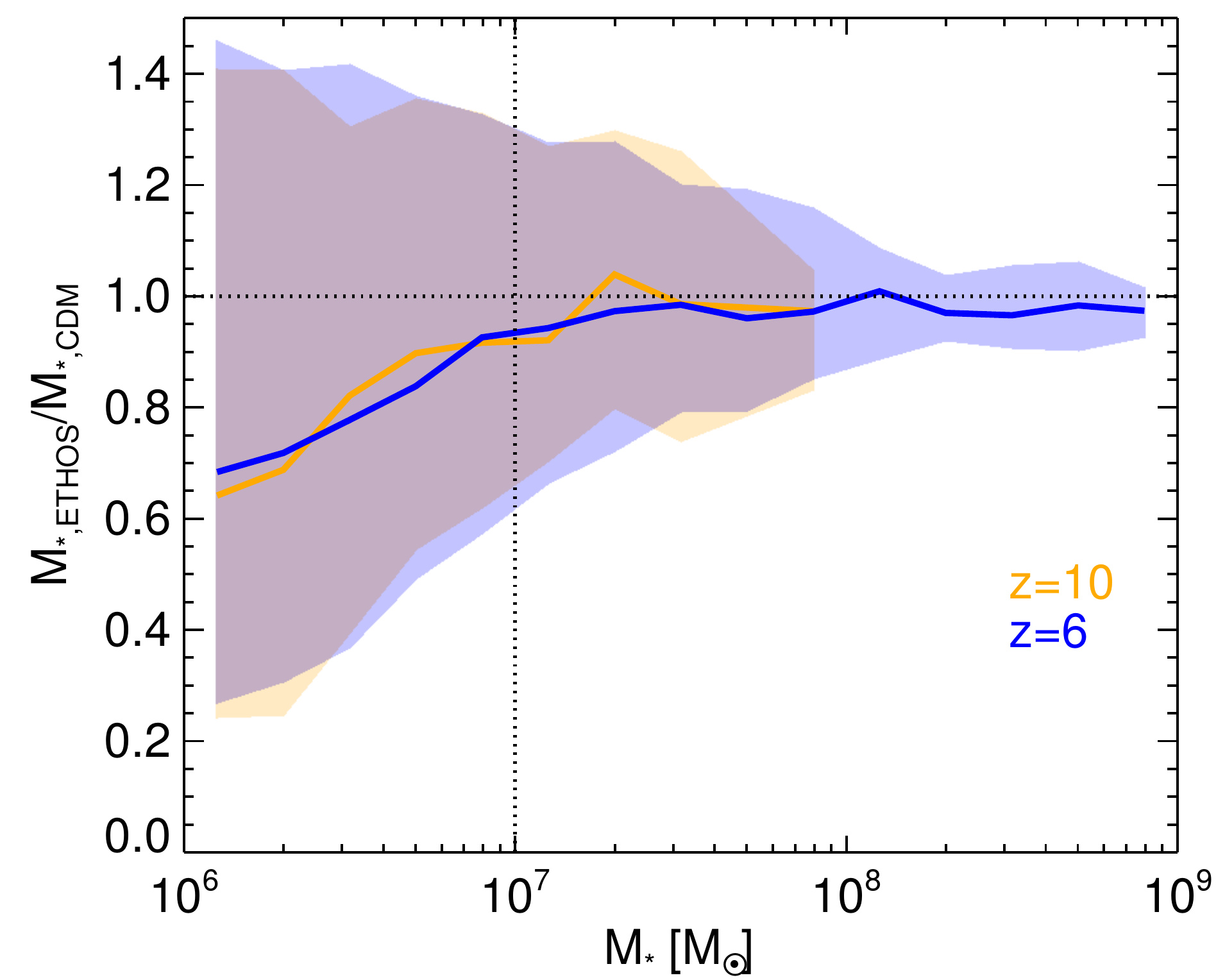}
    \caption{Ratio of stellar masses (ETHOS/CDM) as a function of $M_\ast$ of our bijective matches at $z=6$ (blue) and $z=10$ (orange). The vertical dotted line marks the stellar mass at which each galaxy contains $\sim100$~star particles.}
    \label{MstarMatch}
 \end{figure}

 The stellar mass of each galaxy is one of its most fundamental properties. We would expect based on the previous section that the stellar mass of ETHOS galaxies would be suppressed in ETHOS compared to CDM, if the stellar mass follows $M_{200}$. We examine this hypothesis in Fig.~\ref{MstarMatch}, in which we perform the same analysis as Fig.~\ref{M200Match} but with the stellar mass, $M_{*}$, rather than $M_{200}$; we restrict our sample of matched halo pairs to those haloes that contain star particles in both their CDM and ETHOS matches. From hereon in our definition of stellar mass is the total mass in star particles that is gravitationally bound to the parent dark matter halo.    
At $z=6$, there is a persistent tendency for the median of galaxies with $M_{*}>10^{7}\msun$ -- which corresponds to 100 star particles -- to be suppressed by only a modest 5~per~cent in ETHOS compared to CDM, whereas less massive galaxies are progressively more suppressed in ETHOS, up to 30~per~cent at $10^{6}\msun$. However, these median relations are not tracked by the wider distribution (the upper limit of the central 68~per~cent of the data continues to rise at the lowest masses plotted), and $10^{6}\msun$  corresponds to only $\sim10$ star particles. This result is replicated very closely at $z=10$, with resolved galaxies showing a minimal change in the stellar mass.     
 
\begin{figure}
   \includegraphics[scale=0.45]{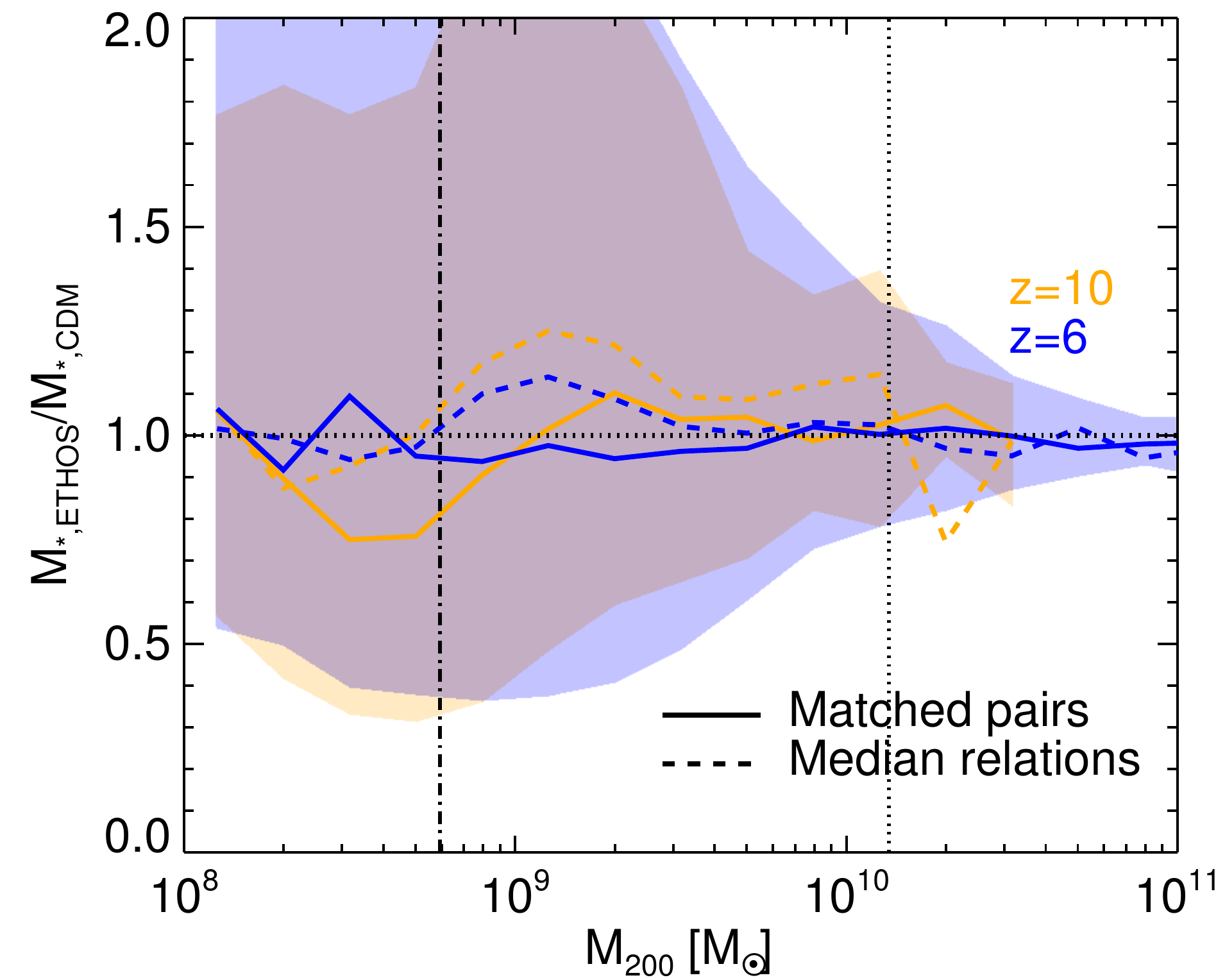}
    \caption{Ratio of the stellar mass-halo mass relation at $z=6$ (blue) and $z=10$ (orange). Solid lines show results for matched pairs (in which case $M_{200}\equiv M_{200}({\rm CDM})$), and dashed the ratio of the CDM and ETHOS median relations (in which case the halo masses are the corresponding ones for CDM and ETHOS). The vertical dotted line marks the approximate $M_{200}$ at which haloes obtain 100 star particles, and the dot-dashed line shows the position of the half-mode mass, $M_\rmn{hm}$, of the ETHOS model.}
    \label{SmHm8}
    \end{figure}

If it is indeed the case that the stellar mass, in so far as it is well resolved, changes little between CDM and ETHOS, this implies a puzzle given that we have already shown that $M_{200}$ certainly does change. We examine this relationship further by repeating the ratio of ETHOS-CDM matched pair stellar mass as a function of the CDM $M_{200}$ and present the results in Fig.~\ref{SmHm8} (solid lines). We also compute the median $M_{*}$ as a function of $M_{200}$ -- the stellar mass-halo mass relation -- for both sets of haloes separately, then compute the ratio of the two medians (this is shown with dashed lines). We only use ETHOS and CDM haloes that have bijective matches, thus the data included in both the median-of-ratios and ratio-of-medians are the same.
 
The size of the scatter in Fig.~\ref{SmHm8} is large, indicating once again the resolution issues at small stellar masses (inevitably in small haloes). The median relations for matched pairs are nevertheless very flat and close to one, despite the fact that the halo mass is clearly suppressed in ETHOS below $10^{10}\msun$ (see Fig.~\ref{M200Match}). We are reminded of this difference in the bump up to 1.1 in the ratio of the two median relations around $10^{9}\msun$ at $z=6$, showing that, since the ETHOS halo mass is lower but the stellar mass is the same, the median stellar mass per unit halo mass is higher than in CDM. An even stronger version of this bump occurs at the same halo mass at $z=10$, up to nearly 30~per~cent above unity. {\it It therefore appears that the star formation efficiency for matched pairs is the same, even though the total mass of ETHOS haloes is lower.} Unlike $z=6$, however, there is a decrement of 20~per~cent in stellar mass between $z=10$ matched pairs at $5\times10^{8}\msun$ (solid orange line): we will return to this point in our discussion of the next Figure.

\begin{figure}
	\includegraphics[scale=0.44]{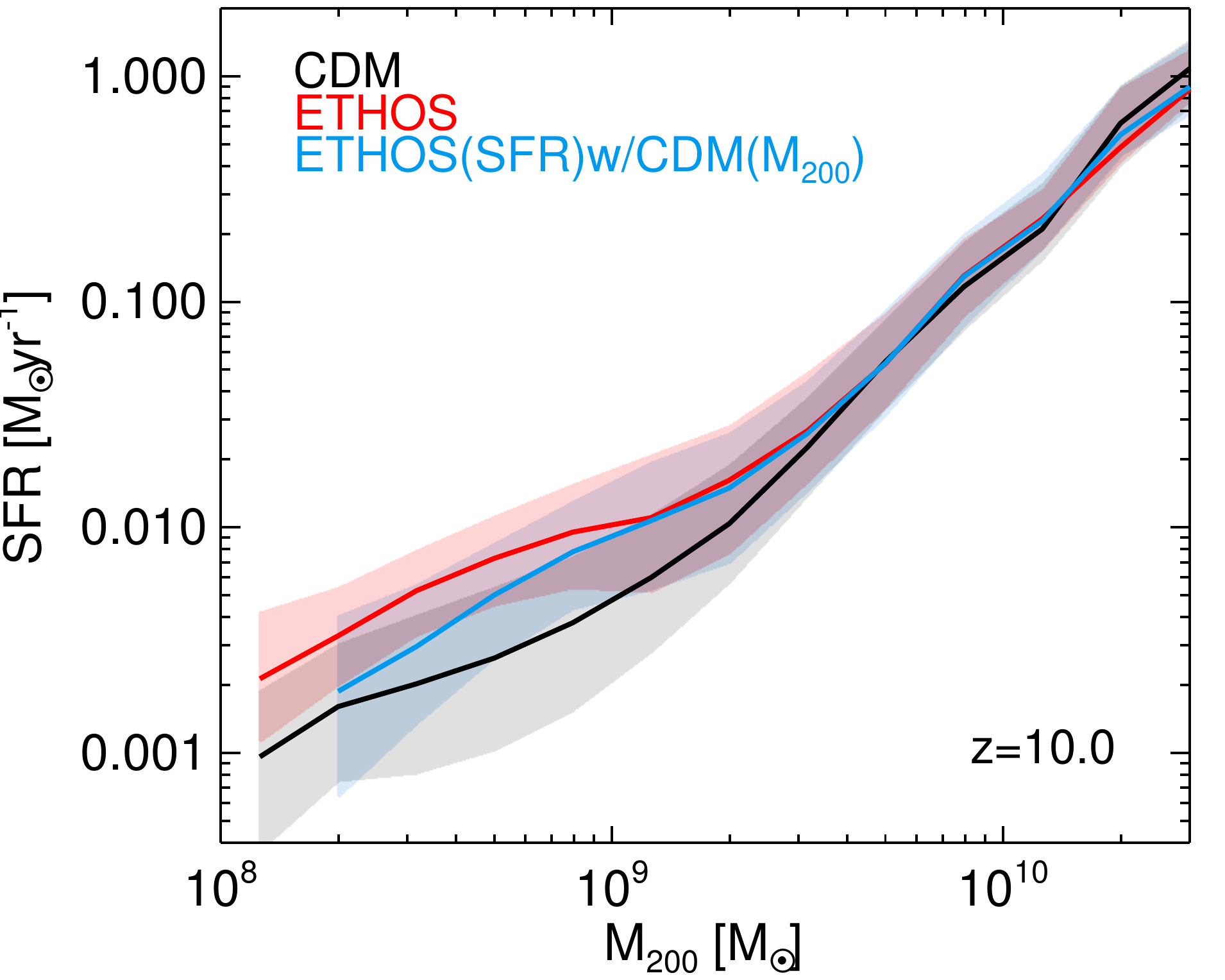}
    \caption{SFR as a function of $M_{200}$ for $z=10$ galaxies in CDM (black) and ETHOS (red). We include a hybrid curve in which the SFR is drawn from the ETHOS galaxies and the $M_{200}$ are measured for the CDM counterpart haloes (blue). The shaded regions encompass $68$~per~cent of the data.}
    \label{SFRMatch}
 \end{figure}

The total stellar mass at different redshifts is inevitably related to the sum of the star formation rate (SFR) at different times. Measuring the star formation rate then has the potential to shed light on why the stellar mass in ETHOS and CDM galaxies is so similar in Fig.~\ref{SmHm8}.
The SFR of each halo is calculated from the sum of the star formation rates of the individual gas cells bound to the halo. 
 Fig.~\ref{SFRMatch} shows the SFRs of matched haloes in ETHOS/CDM as a function of halo mass at $z=10$ (solid lines).
 
Beyond $M_{200}>4\times10^{9}\msun$, there is little difference between models at both redshifts.
but there is a sharp increase in the star formation rate in ETHOS haloes for $M_{200}<4\times10^{9}\msun$. This is consistent with the
scenario suggested in \citet{Bose16c} and \citet{Lovell18a}, where haloes near the cutoff scale in models with a primordial power spectrum cutoff collapse later relative to CDM but do so more rapidly, thus producing a bright starburst. This is shown explicitly with the blue curve in Fig.~\ref{SFRMatch}, which plots the ETHOS galaxy SFR as a function of the counterpart CDM $M_{200}$: we find that the SFR in ETHOS is enhanced by up to a factor of two in ETHOS compared to the CDM counterparts, even though the  halo $M_{200}$ is lower. We have repeated this exercise at $z=6$, and found that at this later time the difference between the models largely disappears. 

  \begin{figure}
   	\includegraphics[scale=0.55]{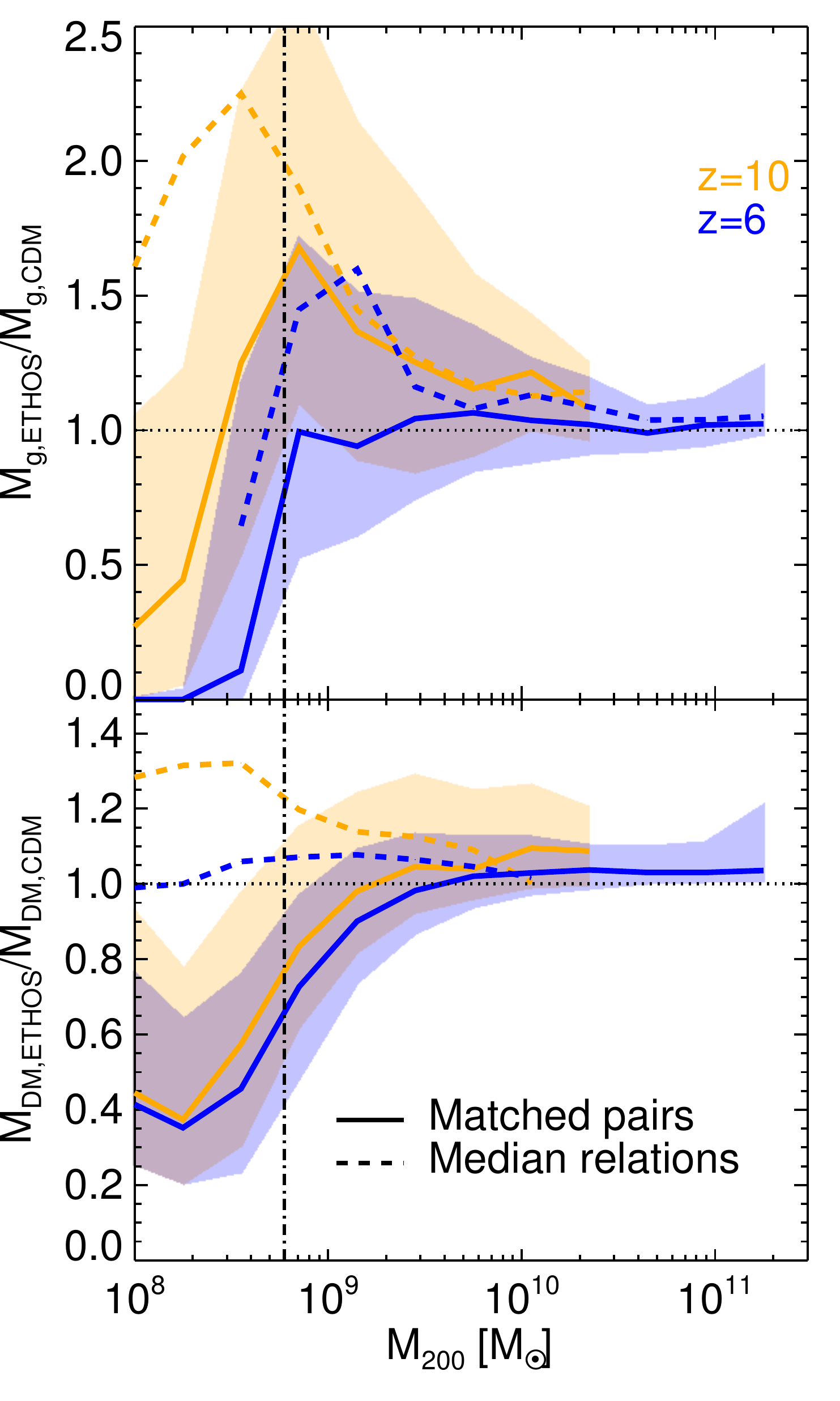}
    \caption{Ratio of the gravitationally bound gas (top panel) and dark matter (bottom panel) masses between CDM and ETHOS galaxies/haloes bijectively matched at $z=10$ and $z=6$ (solid lines). The dashed lines show the ratio of the two median relations in each panel (in which case the halo masses are the corresponding ones for CDM and ETHOS. The vertical dot-dashed line marks the half-mode mass for the ETHOS model.}
    \label{MGaD}
 \end{figure}

The SFR is to a large degree determined by the amount of cold gas in galaxies. This suggests that ETHOS haloes with a higher SFR would have a higher cold gas mass relative to CDM. 
We study this possibility in Fig.~\ref{MGaD}, in which we show with solid lines the relationship between the ratios of gravitationally bound gas (including hot and cold gas, top panel) and dark matter (bottom panel) masses in ETHOS/CDM as a function of virial mass ($M_{200}$(CDM)). As in previous figures, the dashed lines are the ratio of the median relations $M_g-M_{200}$ (top panel) and $M_{\rm DM}-M_{200}$ (bottom panel) for each model.
The gas mass tracks the behaviour of star formation rate (Fig.~\ref{SFRMatch}) closely as expected. We notice that below $M_{200}<M_\rmn{hm}$
evaporation by reionization radiation reduces the amount of gas in the $z=6$ galaxies of both models to very small amounts.

 \begin{figure}
   	\includegraphics[scale=0.45]{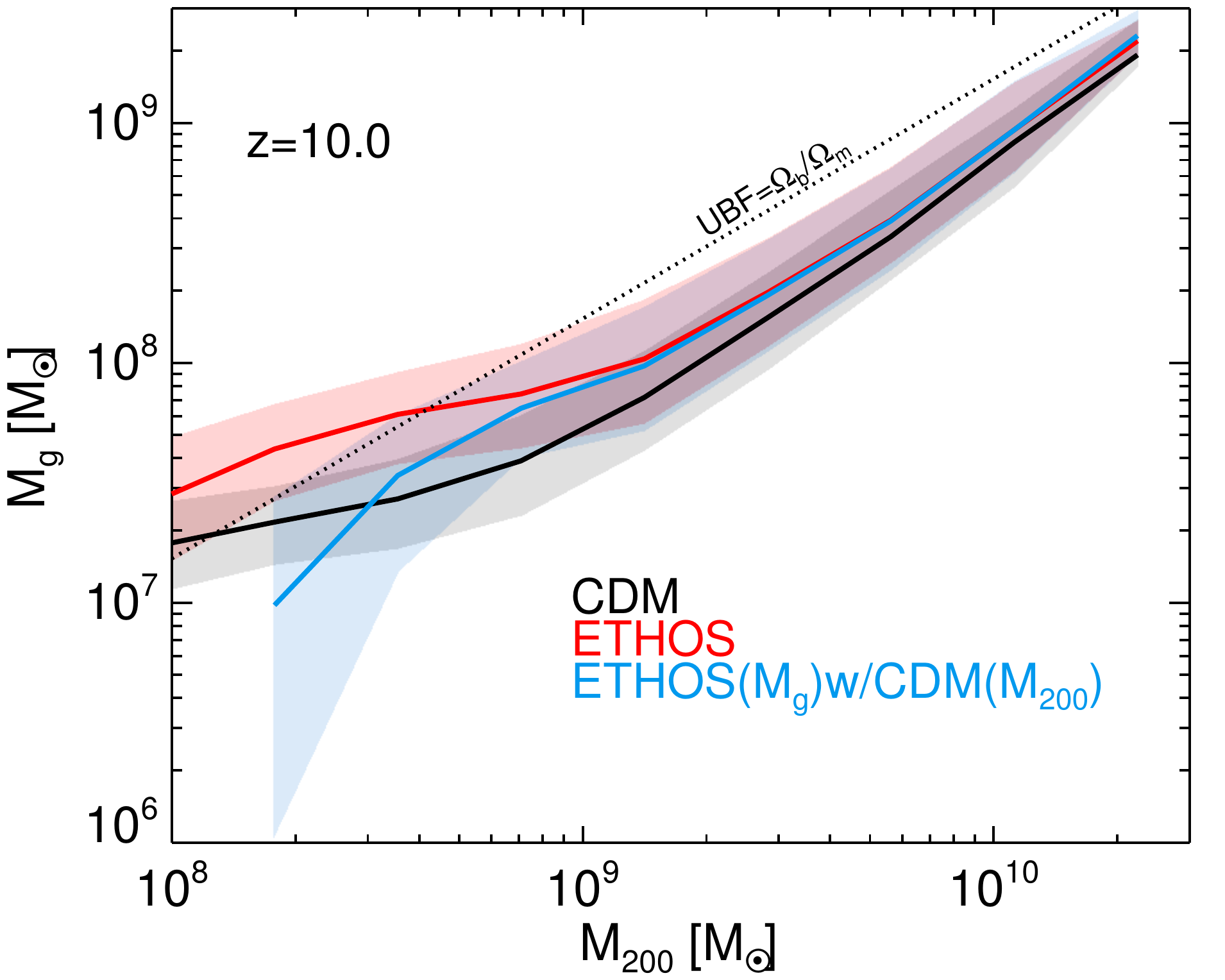}
    \caption{The mass of bound gas in CDM (black) and ETHOS (red) for matched $z=10$ galaxies as a function of $M_{200}$. In blue we show the effect of combining the gas mass measured in ETHOS with the $M_{200}$ measured for the CDM counterpart. The universal baryon fraction is shown as a dotted line.}
    \label{MGas}
 \end{figure}

At $z=10$, the lowest virial-mass ETHOS halo gas masses are suppressed by 70~per~cent relative to CDM, which is to be expected for haloes that have not yet collapsed in ETHOS. The relative gas mass then increases dramatically to 50~per~cent above CDM at the half-mode mass, before returning to CDM values at higher masses. The ratio of the median relations (dashed lines) shows an even higher enhancement in the gas content in ETHOS galaxies over CDM, -- up to 120~per~cent -- for the same reasons discussed in the context of the SFR in Fig.~\ref{SFRMatch}. {\it Overall, the ETHOS haloes have as much gas (if not more) than their CDM counterparts, despite having lower dark matter masses.}

This result is in strong opposition to the status of the dark matter content in haloes, as shown earlier.
This contrast can be seen by looking at the the lower panel of Fig.~\ref{MGaD} (or top panel of Fig.~\ref{M200Match}). Here the dark matter mass of ETHOS haloes is consistently suppressed at the low-mass-end relative to CDM, by 40~per~cent at the half-mode mass, in which the gas mass is at least the same in ETHOS and CDM haloes at $z=6$ and enhanced in ETHOS at $z=10$. The suppression in dark matter mass is stronger at $z=6$ than $z=10$ by about 10~per~cent at most halo masses. It therefore appears that ETHOS haloes have higher baryon fractions: we will return to this point in the next Subsection. Finally, we note that the ratio of median
relations (dashed lines in the bottom panel of Fig.~\ref{MGaD}) implies that ETHOS haloes have {\it higher} bound dark matter masses than CDM haloes of the same virial mass, especially at $z=10$.   

We reinforce this discussion regarding the total gas content of CDM and ETHOS haloes in Fig.~\ref{MGas}, in which we plot the gas mass -- halo mass relations for CDM and ETHOS, separately. We also include a hybrid calculation in which, for our matched haloes, we plot the $z=10$ ETHOS galaxy gas mass as a function of the CDM counterpart halo mass. For $M_{200}<10^9\sun$ the three models are strongly divergent: CDM predicts $2-4\times10^{7}\msun$ of gas in these haloes, ETHOS predicts up to $7\times10^{8}\msun$. The hybrid calculation straddles the two, except at $M_{200}<4\times10^{8}\msun$ where the ETHOS counterparts have not collapsed sufficiently to obtain as much gas as in CDM. We also note that gas masses of ETHOS haloes with $M_{200}<5\times10^{8}\msun$ are above that expected from the universal baryon fraction, notwithstanding the fact that the gas mass is calculated as gravitationally bound mass and $M_{200}$ is defined within a sphere.
 
 \subsubsection{Examples of halo pairs at $z=10$}
 
  \begin{figure*}
 	\includegraphics[scale=0.345]{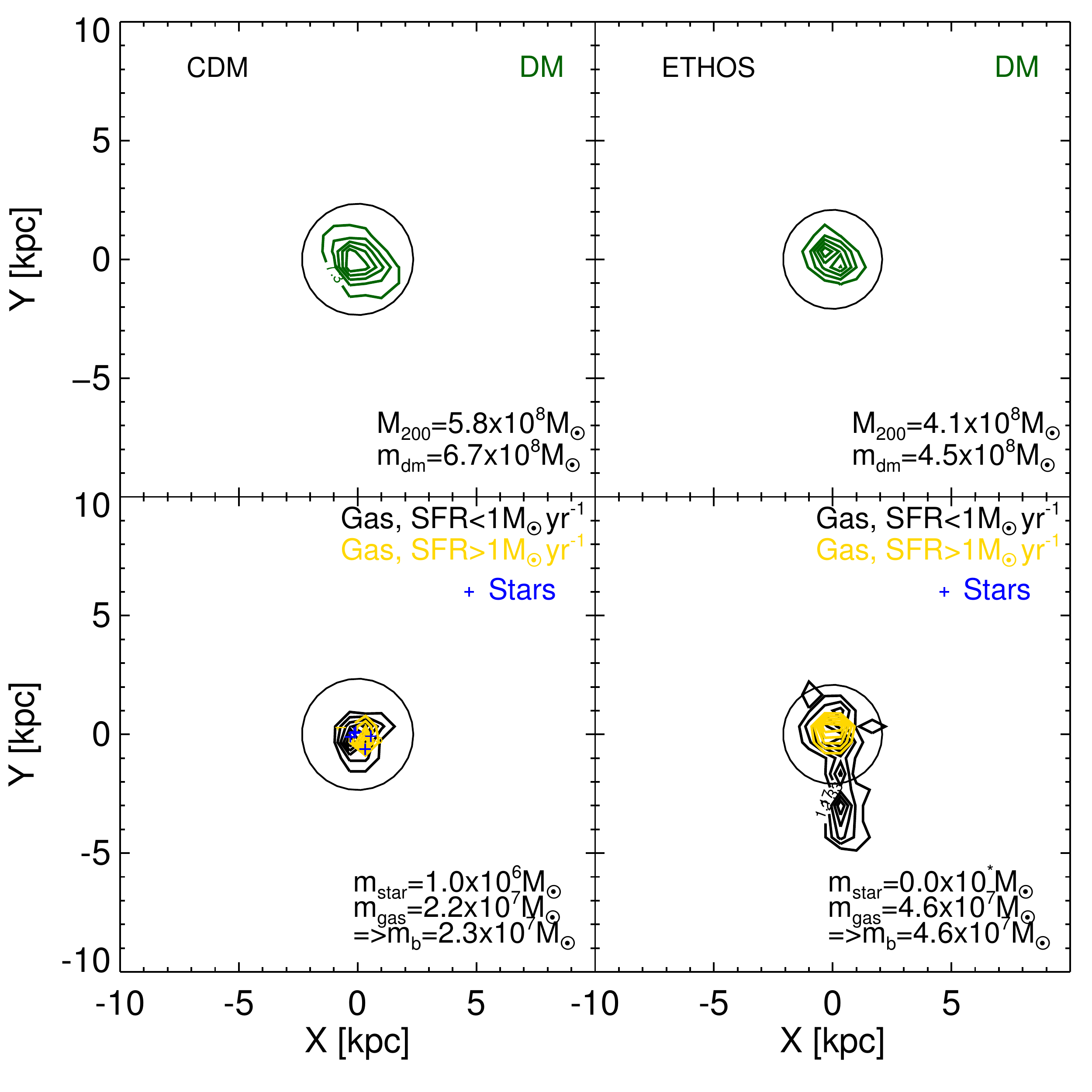}
 	\includegraphics[scale=0.345]{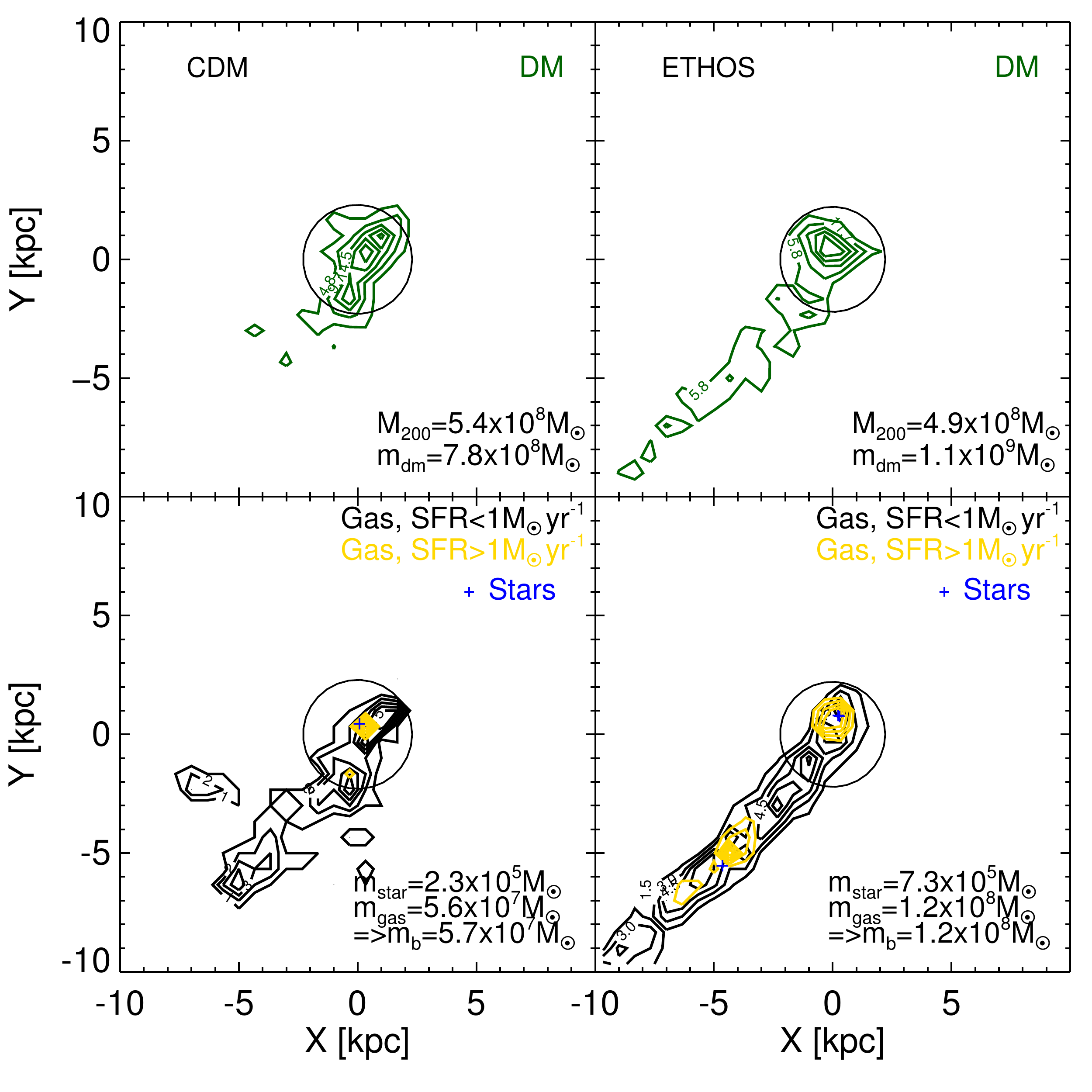}
	\caption{Matter distributions of two examples of matched $z=10$ halo pairs $(R>0.9)$: example i) (left) and example ii) (right). Both halo pairs were selected to have a CDM $M_{200}$ in the range $[5.4,6.0]\times10^{8}\msun$, and the exact values are given in the Figure legends. Each image is 20~kpc (physical) on a side. The top panels show the positions of the dark matter particles as green contours. The bottom two panels show gas cell centroids split by SFR: SFR~$<1\msun\rmn{yr}^{-1}$ (black contours) and SFR~$\ge1\msun\rmn{yr}^{-1}$ (yellow). Star particles are shown in the bottom panel as blue plus signs. For each example pair the CDM halo is shown in the left hand column and ETHOS in the right hand column. In both panels the virial radius, $r_{200}$, is shown as a black circle. Only particles found to be part of the `smooth' central halo by {\sc subfind} are included. The total bound masses in dark matter ($m_\rmn{dm}$), gas ($m_\rmn{gas}$), stars ($m_\rmn{star}$) and baryons ($m_\rmn{b}=m_\rmn{gas}+m_\rmn{stars}$) are given in the Figure legends.}
    \label{Expls}
\end{figure*}
 
 We explore these surprising results -- the excess gas and bound dark matter mass in ETHOS haloes relative to their CDM counterparts -- in the $z=10$ context by taking a few halo examples, and making contour plots of the spatial distribution of matched haloes (comparing also their
virial masses and radii).
We identified 784 high quality ($R>0.95$) matched halo pairs in which the CDM counterpart has a virial mass close to $M_\rmn{hm}$: $M_{200}\in[5.4,6]\times10^{8}\msun$. 
We present two examples of these halo pairs in Fig.~\ref{Expls}. For each halo, we separated their dark matter, gas, and stellar components.   

 The first halo pair example in Fig.~\ref{Expls}, example i) (left panels),  shows the origin of the discrepancy between the varying halo mass definitions and components shown in previous figures. The dark matter distribution in the CDM counterpart is more concentrated than in ETHOS and is more extended spherically. Both distributions show a tail of bound material that extends outside $r_{200}$, towards $Y=-5$~kpc. In the dark matter distributions, this tail has a similar appearance in the two models; however, in the gas distributions this tail is notably denser and more populated in ETHOS. The total gas mass in the ETHOS halo is double that of the CDM counterpart: largely due to the gas tail, but also due to a greater concentration of gas cells within $r_{200}$. The ETHOS halo has yet to form any stars, unlike the CDM counterpart, but has many more highly star-forming gas cells which may be indicative of a coming starburst. We suggest that the early formation of a small stellar population in CDM has already unbound much of the gas through stellar feedback, whereas ETHOS haloes retain their gas until a later point at which either more gas is available for star formation, and/or the gravitational potential is deeper making feedback slightly less effective. A detailed examination of these options will require very fine time resolution, which we defer to a future study. 
 
 The second halo pair example in Fig.~\ref{Expls}, example ii) (right panels), shows a less evolved system in which less of the bound mass has fallen within $r_{200}$. The ETHOS halo shows a particularly striking tail of material that stretches at least 14~kpc from the centre-of-potential. We caution that even the definition of a `halo' is troublesome in this situation \citep[e.g.][]{Angulo13}. The extent of this tail is sufficiently large that the bound dark matter mass is higher for the ETHOS halo than for its CDM counterpart, even though the value of $M_{200}$ is lower in ETHOS. We have calculated the fraction of dark matter and gas mass located outside $r_{200}$, $f(>r_{200})$, for our sample of 784 haloes, and find that for CDM (ETHOS) the median value of $f(>r_{200})$ is 18~per~cent (38~per~cent). For the gas mass, the fractions are higher, at 23~per~cent and 46~per~cent, respectively. This explains the excess bound dark matter mass in ETHOS over CDM at fixed $M_{200}$ shown in Fig.~\ref{MGaD}: the geometry of a smooth, thin, self-bound filament is such that its dark matter mass can greatly exceed its $M_{200}$ unlike the more spherical CDM haloes. For both the CDM and ETHOS haloes the gas tracks the dark matter distribution. The ETHOS halo has twice the gas mass of the CDM counterpart, with an excess of gas in the tail and within $r_{200}$. Unlike in example i), the ETHOS halo has starting forming star particles, yet still has twice as many star-forming gas particles within $r_{200}$. Perhaps even more striking is the presence of star-forming gas -- and even a star particle -- in the tail where there is no hint of a dark matter overdensity. This may be an example of star formation in filaments in dark matter models with a galactic-scale cutoff in the power spectrum as described by \citet{Gao07} in the context of a WDM cosmology. We had stated that this mechanism was rare in our previous paper -- \citep{Lovell18a} -- as we defined filamentary star formation as that which occurs in gas cells not bound to any substructure. The possibility that such star formation occurs in filaments that are in fact bound is an interesting development.
 
 We are left with the puzzle of why the dark matter distribution, and to a lesser extent gas distribution, in young ETHOS haloes will nevertheless evolve into haloes that are less massive, both in $M_{200}$ and in bound mass, than in CDM as expected from example i) (left panel of Fig.~\ref{Expls}) and from Fig.~\ref{MGaD}. The presence of spurious structures from filamentary fragmentation would contribute to $M_{200}$, yet $M_{200}$ is still smaller in ETHOS. We speculate that ultimately much of the material in the long ETHOS halo tails becomes unbound from the central halo, and will either remain unbound until it can accrete into a halo with $M_{200}\gg M_\rmn{hm}$ or may even continue to be unbound up until the present day, as part of the smooth (unclustered) cosmic web. At $z=6$ the percentage of dark matter particles bound to dark matter haloes of any virial mass -- including the mass in centrals and their subhaloes together -- in CDM is 11~per~cent, compared to 7~per~cent for ETHOS, thus in agreement with this hypothesis at least in so far as the bound mass percentage is lower: we will defer a detailed comparison to future work in which the particle distribution is tracked to redshifts $z\ll6$.

  \subsubsection{Condensation times in ETHOS haloes}
\label{sec:ConTimes}
 
\begin{figure}
 	\includegraphics[scale=0.34]{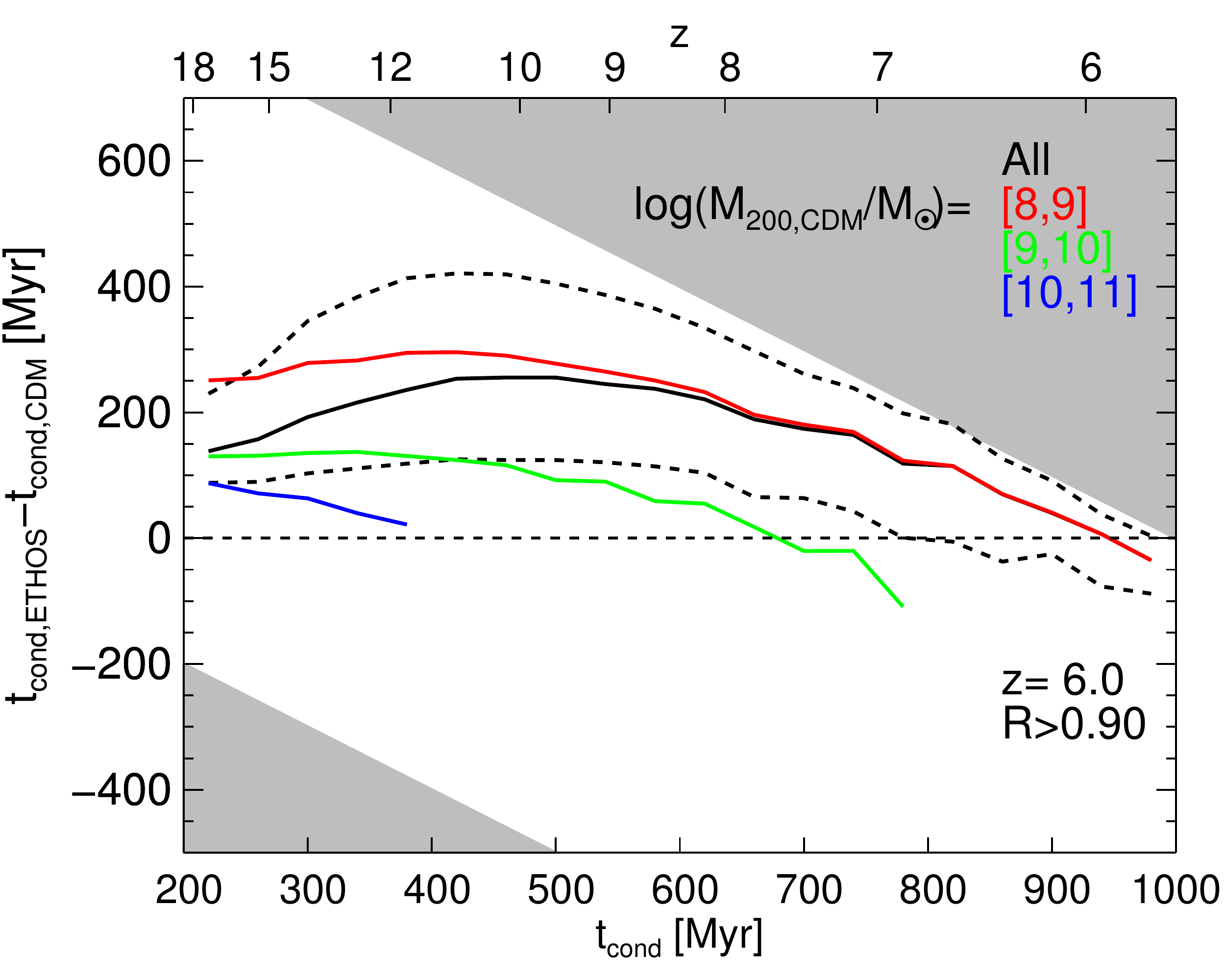}\\
    \caption{Difference between the condensation (cooling) time in ETHOS and CDM haloes    in our bijective matches as a function of condensation time in CDM. The median relation for all halo pairs with $M_{200}(z=6,\rmn{CDM})>10^{8}\msun$ is shown as a solid black line, and its 68~per~cent data region is delineated with dashed lines. We also include the median relations for three bins in $\log(M_{200}(z=6,\rmn{CDM})/\msun)$: [8,9] (red), [9,10] (green) and [10,11] (blue). Regions for which the difference in time is either incompatible with the start of the simulation at $z=127$ or the end of the simulation at $z=6$ are shown in grey.}
    \label{TimeDiffMatch}
 \end{figure}

 While the exact process of how galaxies form in models with primordial power spectrum cutoffs is not entirely clear, it is well known that halo collapse occurs later than in CDM. We show this explicitly in Fig.~\ref{TimeDiffMatch}, in which we plot the difference in condensation time between CDM and ETHOS matched pairs as a function of the condensation time in CDM. The halo condensation time is calculated as the first time at which the halo mass is equal to the mass required for atomic cooling given the halo virial temperature. In practice, since we have discrete time resolution, we interpolate between snapshots to find this time. We only include haloes that achieve high quality bijective matches ($R>0.9$) and $M_{200}>1\times10^{8}\msun$ in both CDM and ETHOS .     
 
\begin{figure}
 	\includegraphics[scale=0.66]{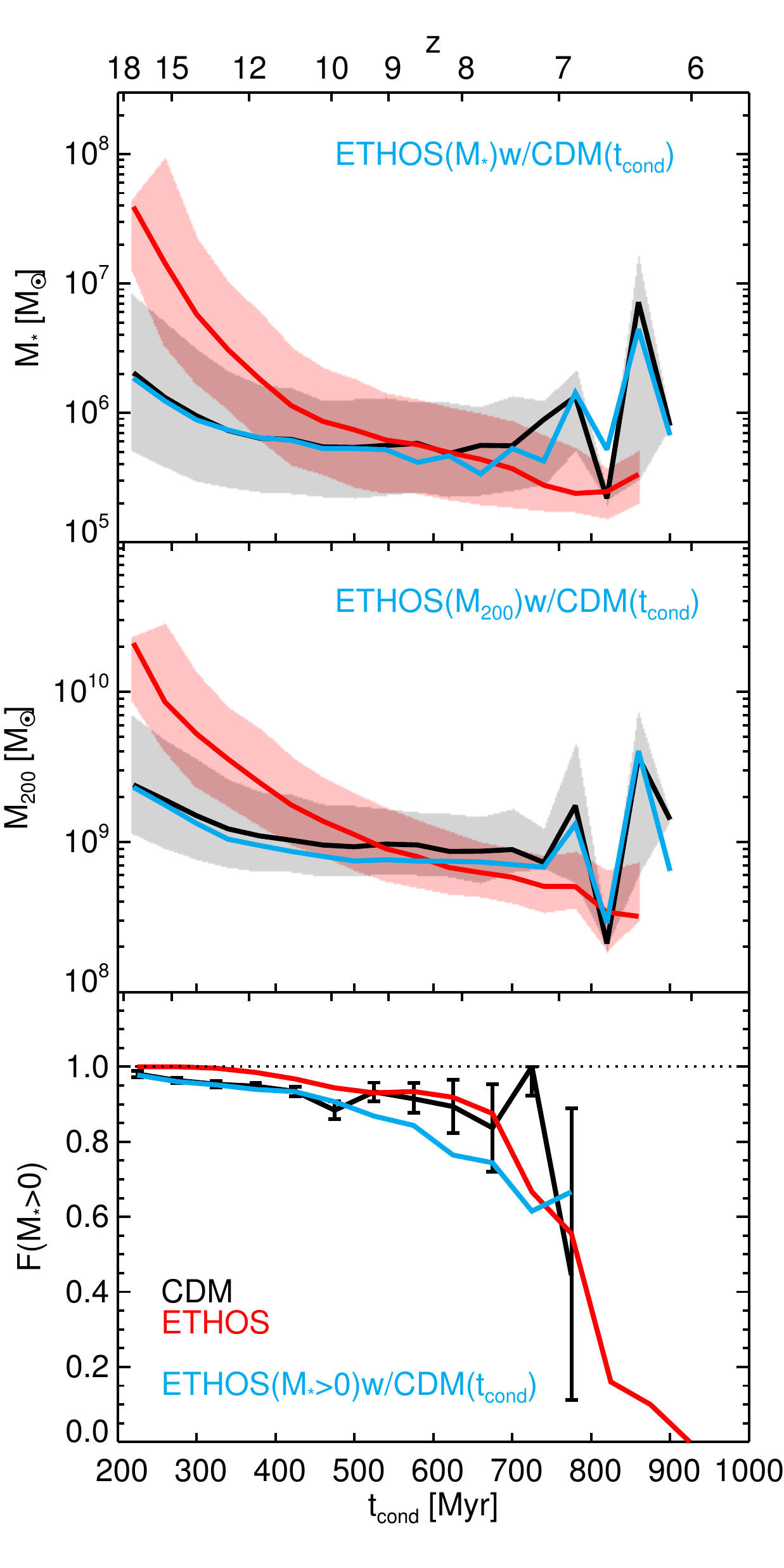}\\
    \caption{Median halo properties of $z=6$ galaxies as a function of their condensation time. Top panel: stellar mass. Middle panel: halo mass. Bottom panel: luminous fraction (`luminous' = contains at least one star particle), in which error bars ($2\sigma$ binomial) are shown only for CDM (for clarity purposes). The black lines show results for CDM haloes (in the matched pairs set) and the red for the corresponding ETHOS haloes. The blue lines show the results for ETHOS haloes but using the matched CDM halo condensation times. We include the redshifts corresponding to each condensation time on the upper $x$-axis.}
    \label{TimeDiffMatch1}
 \end{figure}

 The ETHOS haloes on average condense 100 to 200~Myr later than their CDM counterparts. The difference between condensation times actually grows slightly from haloes that in CDM condense at $\sim$200~Myr to those at $\sim$400~Myr, but then the curve turns over afterwards. We suspect that this turn over occurs because the condensation delay for ETHOS haloes with CDM counterpart that condense at times $>400$~Myr is too long to occur by $z=6$ (see grey regions in Fig.~\ref{TimeDiffMatch}). Also, we note that the vast majority of CDM haloes condense before 400~Myr, and therefore there is the possibility of being affected by sampling noise at later times (see the error bars in the bottom panel of Fig.~\ref{TimeDiffMatch1}.)    
 
 This delay in collapse time has been shown to affect the inner densities ($<1$~kpc) for a small population low mass objects at $z=0$ in dark matter-only simulations \citep{Lovell12}. We take advantage of the large sample of hydrodynamical galaxies here, coupled with our matching algorithm, to examine what effect these later collapse times, and thus condensation times, in ETHOS have on the total halo mass and stellar mass --  including the galaxy formation efficiency -- of these haloes. In other words, is there an additional effect on the main galaxy properties due to this delay in the condensation time, or is it simply a reset of the onset of galaxy formation? 

 We approach this question in the following manner. We first calculate the median stellar mass and halo mass as a function of condensation time for our $z=6$ ETHOS and CDM matched-pair haloes, treating ETHOS and CDM independently. We then compute the same median relations for ETHOS haloes as a function of their CDM counterpart condensation times. If the latter hybrid calculation tracks the CDM median relations then the delayed condensation time in ETHOS has not had an effect on the evolution of the relevant galaxy property, it is simply a reset of the onset of galaxy formation given by the primordial power spectrum cut-off. We present the results in Fig.~\ref{TimeDiffMatch1}. In the stellar mass and halo mass plots we only include haloes that have formed at least one star particle in both the ETHOS and CDM matched counterparts. We consider whether this approach is reasonable by also plotting the luminous fraction as a function of condensation time in the same Figure (bottom panel). 
 
   \begin{figure*}
 	\includegraphics[scale=0.68]{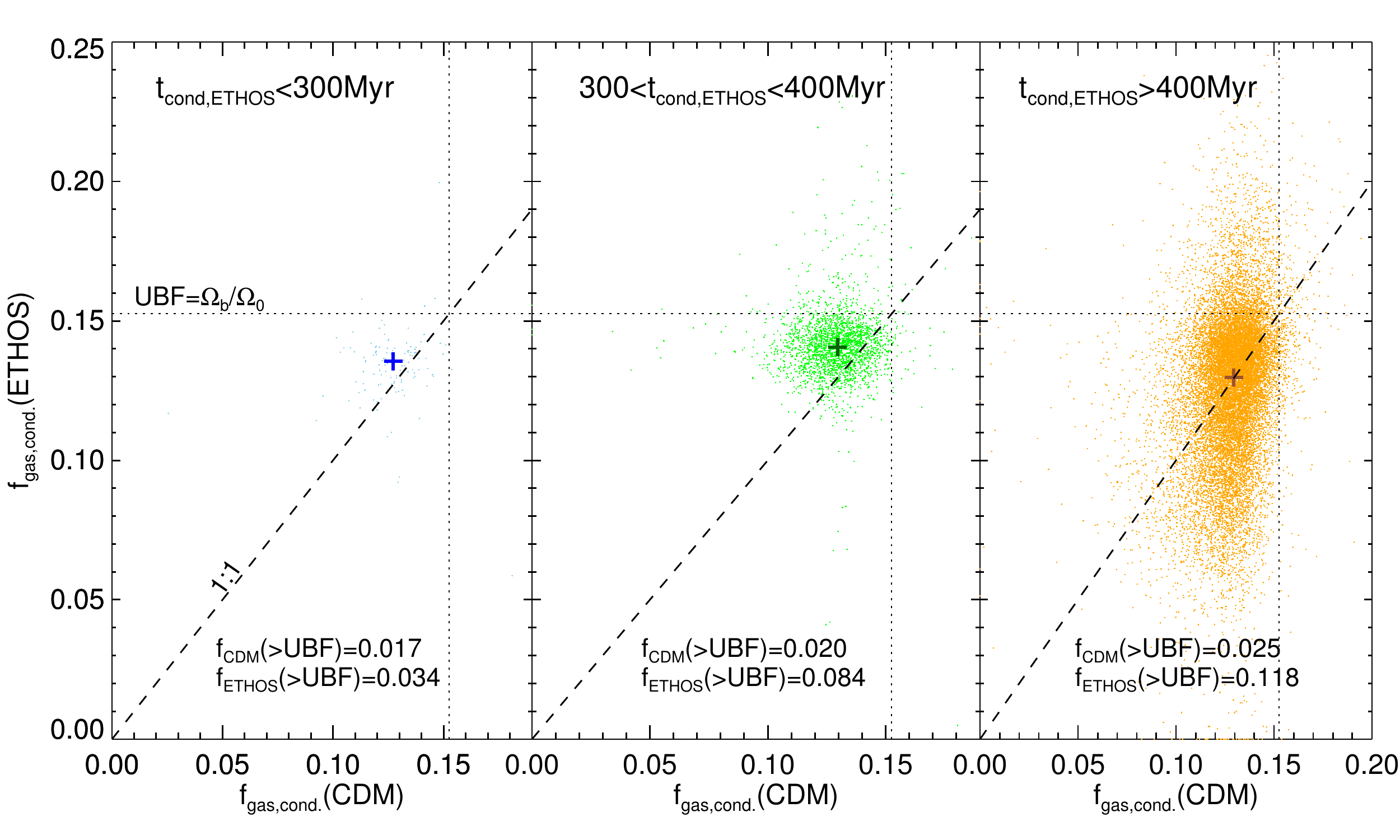}
    \caption{Gas fractions of CDM-ETHOS halo pairs in three bins of ETHOS halo condensation times, $t_\rmn{cond,ETHOS}$: $t_\rmn{cond,ETHOS}<300$~Myr (left panel), $300<t_\rmn{cond,ETHOS}<400$~Myr (middle panel) and $400<t_\rmn{cond,ETHOS}$ (right panel). Alternatively, these panels corresponds approximately to ETHOS condensation redshifts $z_{\rm cond}>12$, $12>z_{\rm cond}>8.5$ and $z_{\rm cond}<8.5$ respectively. We plot the CDM halo gas fractions on the x-axis and their ETHOS counterpart gas fractions on the y-axis. The one-to-one relation is shown in each panel as a diagonal dashed line, and the universal baryon fraction as a pair of dotted lines. The centre of each distribution, as defined by the medians of the two gas fractions, is indicated by a plus sign. The fraction of galaxies that have gas fractions higher than the universal baryon fraction -- $f(>\rmn{UBF})$ -- for each dark matter model / ETHOS condensation time combination is indicated for CDM and ETHOS in the panel legends.}
    \label{GasFrac}
 \end{figure*}

 At face value, the CDM and ETHOS models paint a very different picture of both the stellar mass and halo mass formation histories as a function of condensation time. The ETHOS haloes that have a condensation time of 200 Myr ($z_{\rm cond}=18$) have a median $z=6$ stellar mass of $4\times10^{7}\msun$, more than an order of magnitude higher than CDM. The ETHOS $M_\ast-t_{\rm cond}$ relation then declines much more rapidly than CDM, and the two meet at 600~Myr for a median $M_{*}=6\times10^{5}$, which is only 6 star particles. However, our hybrid calculation, with the ETHOS-$M_{*}$ coupled to the counterpart CDM-$t_{\rm cond}$, instead tracks the CDM relation almost perfectly, even in the region around $z_{\rm cond}=7$ where the star particle sampling is very poor. We therefore recover one of the results from Fig.~\ref{MstarMatch}, namely that the stellar masses of matched pairs are very similar and are not affected by the delay in condensation time. 

 We next consider the connection between halo mass and condensation time in the middle panel of Fig.~\ref{TimeDiffMatch1}. The ETHOS haloes that form at the earliest times are more massive than in the CDM relation, and have a $M_{\rm 200}-t_{\rm cond}$ relation that falls off more steeply towards larger condensation times converging with CDM by $z_{\rm cond}=6$. The hybrid median relation shows agreement with CDM at the earliest times, but unlike the case of the stellar mass (top panel of Fig.~\ref{TimeDiffMatch1}) peels off towards lower redshifts so that the median halo is almost 10~per~cent less massive than CDM by redshift $z_{\rm cond}=6$. This is because haloes that form later have had less time to collapse in ETHOS. We have repeated this exercise with all matched haloes independently of whether they host any star particles, and find a much bigger suppression, by up to 50~per~cent.
 
 We have previously shown that the stellar masses do not change markedly between CDM and ETHOS, and that any changes that are present at $z=10$ are swiftly erased by higher subsequent SFRs in ETHOS. However, we restricted that analysis to haloes that form stars in {\it both} CDM and ETHOS. We therefore examine the situation with haloes for which this is not necessarily true. We plot the luminous fraction of haloes -- haloes with at least one star particle -- with a $z=6$ halo mass $>10^{9}\msun$ ($>10^{8}\msun$) in CDM (ETHOS) as a function of their condensation time in the bottom panel of Fig.~\ref{TimeDiffMatch1}; we also restrict ourselves to matches with $R>0.9$. Note that this is only the luminous fraction of bijective matched haloes: we do not take into account the greater total abundance of CDM haloes. The ETHOS luminous fractions are consistently higher than those in CDM for $z_{\rm cond}>10$, and at lower redshifts agree more with CDM although the (binomial) error bars in this regime are very large due to the small number of late-forming CDM haloes. The hybrid calculation tracks CDM down to $z_{\rm cond}=10$ and then exhibits a lower luminous fraction from then on. This result suggests that $z=10$ is the redshift at which reionisation feedback starts to become important; prior to this redshift the choice of dark matter model has no effect on the luminous fraction.  

 We have shown that the degree to which the properties of galaxies are influenced by the different condensation times of the haloes in ETHOS and CDM is is simply driven by a delay of the onset of galaxy formation in the former caused by the primordial power spectrum cutoff. In particular, we showed how star-forming CDM haloes mostly condense above $z=10$ irrespective of mass whereas the ETHOS counterparts experience a delay in condensation such that longer delays occur for less massive haloes. We now make the link from the differences in condensation time shown in Fig.~\ref{TimeDiffMatch} to the changes in the SFR shown in Fig.~\ref{SFRMatch}, as mediated by the gas fraction at the condensation time: higher gas fractions, particularly in low mass haloes from which hot gas is readily ejected by supernovae, will be expected to have higher SFRs. We separate out our matched pairs ($R>0.9$) into three bins in ETHOS condensation time $t_\rmn{cond,ETHOS}$: $t_\rmn{cond,ETHOS}<300$~Myr, $300<t_\rmn{cond,ETHOS}<400$~Myr and $400<t_\rmn{cond,ETHOS}$, and for each sample plot the ETHOS gas fractions as a function of the CDM counterpart gas fractions in Fig.~\ref{GasFrac}. We restrict ourselves to pairs in which the ETHOS haloes have the following properties at $z=6$: $M_{200}>10^{8}\msun$, and at least one star particle.    
 
 All three subsamples of halo pairs show similar distributions of CDM gas fractions -- centred on $\sim0.13$ -- which is unsurprising since we have already shown that all of the CDM haloes condense at roughly the same time ($z_{\rm cond}>9$) and therefore largely have the same opportunity to expel gas through stellar feedback. The ETHOS galaxies on the other hand, show some evolution with time, from a median $f_\rmn{gas}=0.14$ in the earliest bin down to $0.13$ in the latest bin. This effect is likely linked to the additional reionisation feedback that partially evaporates the gas in later forming haloes, and may also reflect extra energy from nearby supernovae. 
 
Although the gas fraction of ETHOS haloes evolves more strongly with redshift than their CDM counterparts, we find that the ETHOS gas fractions at the earliest times are significantly higher. In the median this is 1~per~cent (absolute percentage), but we also find many haloes with gas fractions higher than even the universal baryon fraction (UBF). On average, approximately 10~per~cent of matched ETHOS haloes across all times have a gas fraction above the UBF compared to only $\sim2$~per~cent of CDM haloes. One possibility for this difference is the way that halo finders sometimes incorporate filaments into haloes \citep[see e.g.][]{Angulo13}, but this effect is observed for the friends-of-friends algorithm rather than for an algorithm based in gravitational binding energy as the one we use here. Another, plausible explanation is that this excess of gas in ETHOS is due to monolithic collapse of the first generation of star-forming haloes near the cutoff scale. Unlike in CDM, in ETHOS, there has been no previous generation of lower mass haloes in which gas can start forming stars prior to the model condensation time, and perhaps collisional cooling in the cosmic web is able to funnel extra gas into the halo at super-UBF levels.  In any case, we have shown that ETHOS haloes retain larger fractions of their gas than CDM at the condensation time. This provides an explanation for the spike in ETHOS SFRs presented in Fig.~\ref{SFRMatch}, which in turn explains how on a halo-by-halo basis ETHOS haloes have stellar masses broadly similar to their CDM matched counterparts.
 
   \subsubsection{The onset of star formation in ETHOS}
 
We have hinted so far at a potentially interesting difference between the first galaxies in CDM and ETHOS: CDM haloes start to form stars earlier than their ETHOS counterparts, and given the lower stellar masses of ETHOS galaxies at $z=10$, and the faster build up of stellar mass to catch up with CDM by $z=6$, we therefore anticipate an older stellar population in CDM than ETHOS {\it at a fixed stellar mass}. We end this section by looking at the rapidity of the star formation history in the two models and also look ahead to possible observational signatures.
 
\begin{figure}
 	\includegraphics[scale=0.34]{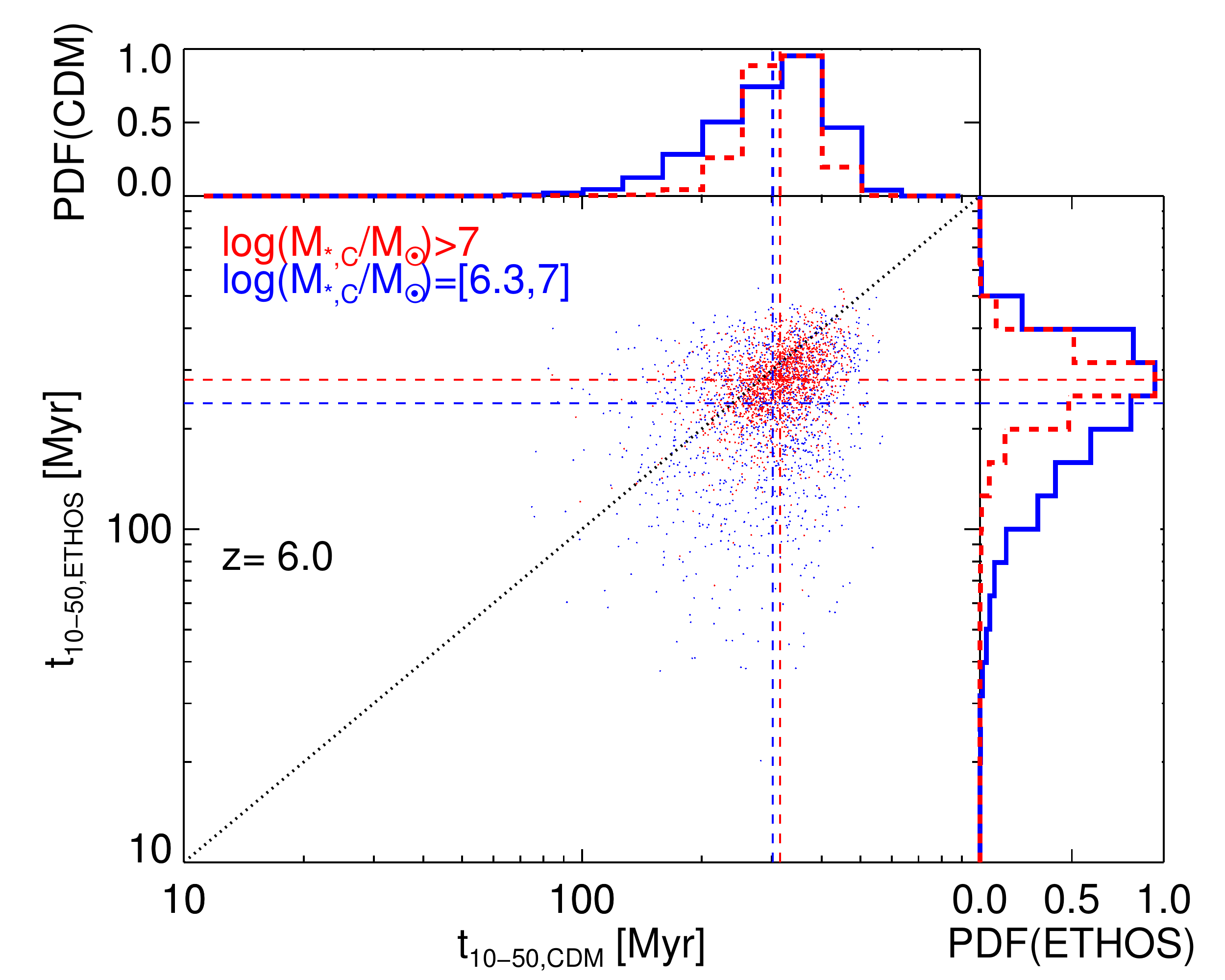}
    \caption{Time taken to form the first 50~per~cent of the stellar mass by z=6 in our bijective matches, with ETHOS values on the $y$-axis and CDM values on the $x$-axis. The PDFs of these quantities are shown in the insets, and are normalized to the maximum value of the histogram to improve legibility. We present the results for two stellar mass ranges, $M_\rmn{*,CDM}>10^{7}\msun$ (red points) and $10^{6.3}<M_\rmn{*,CDM}<10^{7}\msun$ (blue points). }    
    \label{T50pcMatch}
 \end{figure}

 We measure the speed of the build up of the stellar mass by calculating the time it takes for each galaxy to form the first 50~per~cent of the total amount of stars at $z=6$. In practice, we extract the distribution of ages of the stars that constitute each galaxy at $z=6$, calculate the median and subtract the first decile of that distribution. In other words, if the first 50~per~cent of the stellar mass is formed by time $t_{50}$ and the first 10~per~cent by time $t_{10}$, the time that we measure is $t_{50}-t_{10}$, or $t_{10-50}$. We calculate this difference for all of our matched pairs, and plot the results in Fig.~\ref{T50pcMatch} with the ETHOS  $t_{10-50}$ as a function of the matched CDM  $t_{10-50}$. We split the distribution into two subsamples based on resolution: well resolved $M_\rmn{*,CDM}>10^{7}\msun$ (red in Fig.~\ref{T50pcMatch}) and more tenuously resolved, $5\times10^{6}<M_\rmn{*,CDM}<10^{7}\msun$ (blue points) 

 The locus of the higher stellar mass galaxies is at [315,280]~Myr (medians), showing that ETHOS galaxies generally form 30~Myr faster than their CDM counterparts; for the lower mass galaxies shown the difference grows to 60~Myr. It is also interesting to note that only $12$ ($<1$~per~cent) of the CDM haloes -- mostly in the marginally resolved sample -- generate their mass in under 100~Myr compared to $\sim100$ (8~per~cent) of the ETHOS haloes. We therefore conclude that 
a modest survey conducted at $z=6$, with observational estimates of how fast the stellar mass was assembled, might constitute evidence against or in favour of currently allowed models that have a primordial power spectrum cutoff.  
 
\begin{figure}
 	\includegraphics[scale=0.34]{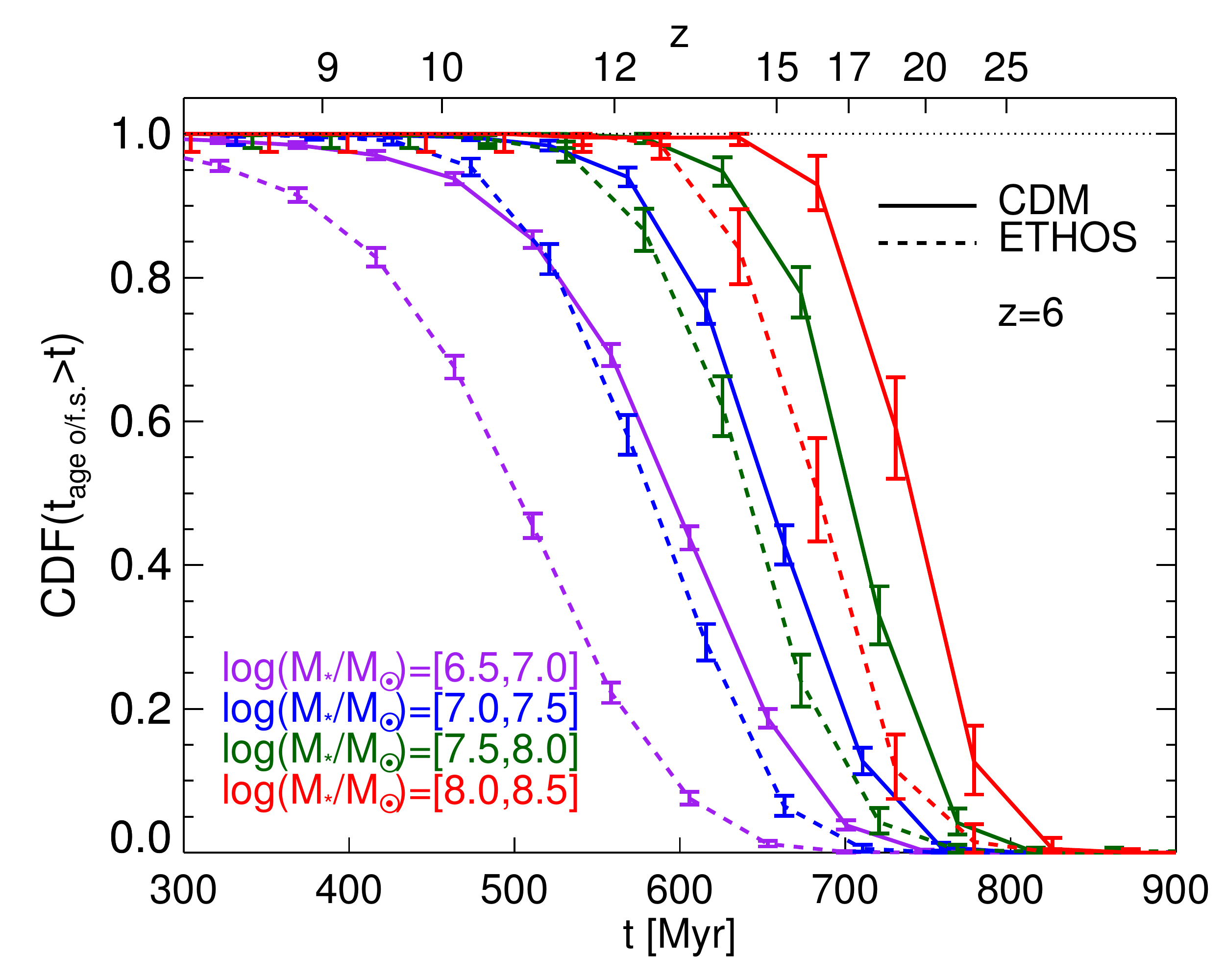}\\

    \caption{PDF of what proportion of $z=6$ galaxies have at least one star particle (given our resolution, each star particle has a mass $\sim10^{5}\msun$) older than time $t$ in four logarithmic stellar mass bins as indicated in the Figure legend. The CDM (ETHOS) case is shown with solid (dashed) lines. Error bars show $2\sigma$ binomial errors.}
    \label{AgePDF}
 \end{figure}

Attempts to detect a difference between these models in this manner might become feasible thanks to high quality SEDs of galaxies at $z>6$. Recently, \citet{Hashimoto18} examined the SED of a lensed galaxy at $z=9$ and reported the detection of a stellar population that formed $10^{9}\msun$ in stars at $z\sim15$. This sort of work has the potential to favour ETHOS if the formation time is sufficiently rapid, as shown in the previous Figure, or alternatively to rule out ETHOS if the first stars in a given population of galaxies form at a redshift before any ETHOS structures have collapsed. In regards to the latter, we therefore compute the PDF of how many $z=6$ galaxies have at least one star particle older than a given time in four stellar mass bins ($\log{(M_{*}/\msun)}=[6.5,7.0],[7.0,7.5],[7.5,8.0],[8.0,8.5]$) for all CDM and ETHOS galaxies independently, regardless of whether they have a match in the other simulation. We plot the results in Fig.~\ref{AgePDF}  

 For all stellar mass bins there is a statistically significant difference between the proportion of CDM and ETHOS galaxies that have at least $\sim10^{5}\msun$ (one star particle) in stars older than a given age threshold. The difference is smallest for the highest stellar mass bin, since these objects form early in both CDM and ETHOS, and is largest for the smallest mass bin, where the delay in ETHOS halo collapse is longest. We find that all models and stellar masses successfully produced some galaxies with stellar populations that formed around $z=15$ as in \citet{Hashimoto18}, although for the lowest mass bin in ETHOS this is only 5~per~cent of galaxies. The ideal target for an observational comparison would be the lowest mass bin given the good statistics, but even the more observationally accessible bin ([8.0,8.5]) holds some promise. We predict that if fewer than 50~per~cent of bright galaxies ($\log{(M_{*}/\msun)}=[8.0,8.5]$) observed at $z=6$ have stellar populations older than 700~Myr, this would provide indirect evidence for the presence of a cutoff in the linear matter power spectrum. We expect that the only way the effect could be mimicked in CDM is through very efficient feedback from the first stars.
 
Next, we consider the possible implications for dwarf galaxies in the Local Universe, where it is possible to observe faint galaxies at the threshold of galaxy formation, and thus where the difference between models such as ETHOS and CDM becomes more distinct. The best link between high redshift galaxies and Local Group dwarfs are likely `fossil galaxies' of the kind described by \citet{Bose18}, which were quenched by reionization feedback (photo-heating evaporation) 
in the run up to $z=6$. We therefore select CDM and ETHOS galaxies that are gas poor at $z=6$, which we define as having a gas mass lower than 10~per~cent of the stellar mass, and repeat the same process as for Fig.~\ref{AgePDF} to study the differences in the age distribution. We only include galaxies in the lowest stellar mass bin, $\log{(M_{*}/\msun)}=[6.5,7.0]$, due to the lack of gas-poor galaxies in the other bins.     
 
  \begin{figure}
 	\includegraphics[scale=0.34]{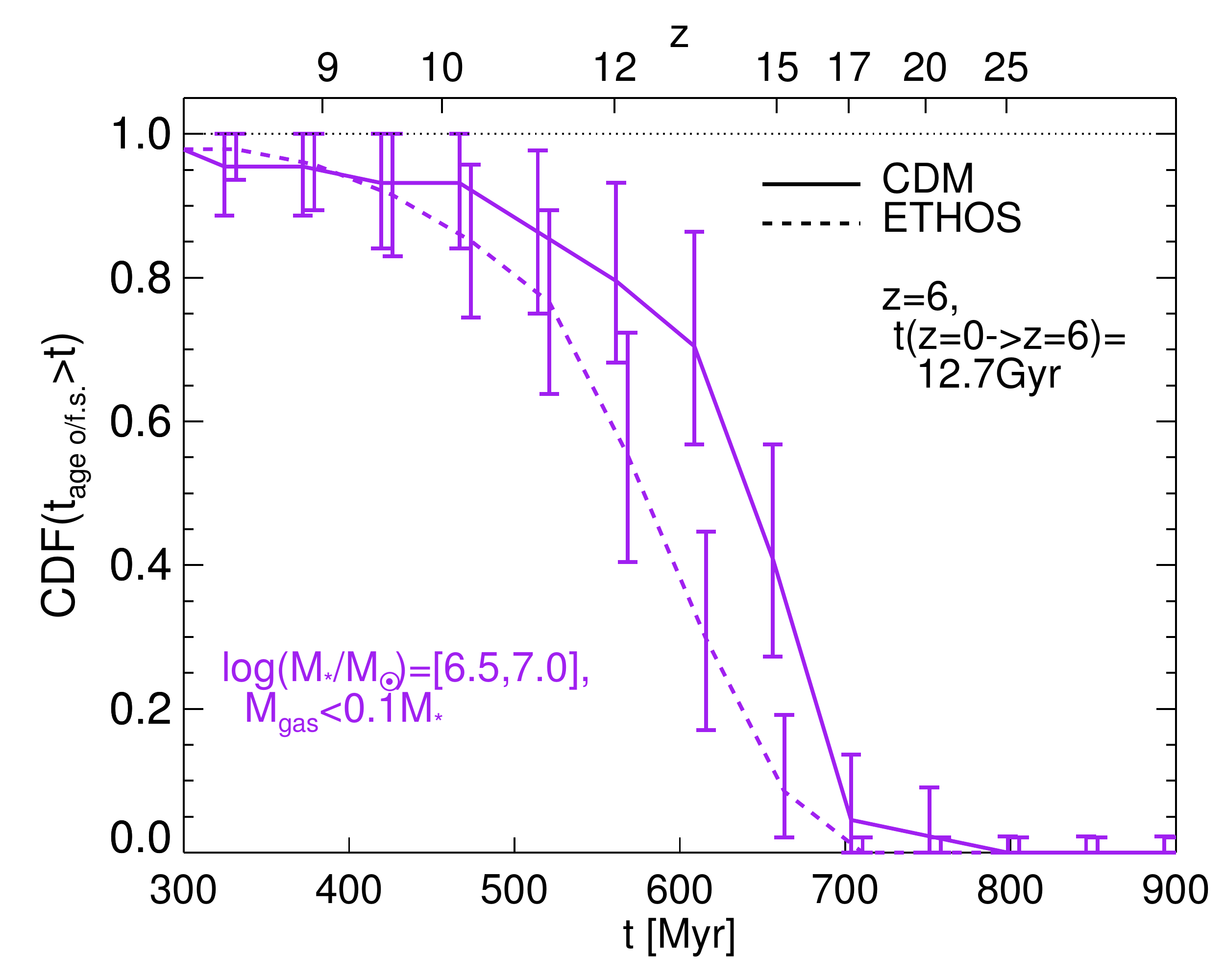}\\

    \caption{PDF of the ages of the oldest star particles in CDM and ETHOS galaxies whose gas mass is less than ten~per~cent of their stellar mass at $z=6$. We only include galaxies from the lowest mass bin in Fig.~\ref{AgePDF}, $\log{(M_{*}/\msun)}=[6.5,7.0]$, as the other mass bins do not contain enough gas-poor galaxies to see a statistically significant difference between CDM and ETHOS. Error bars show $2\sigma$ binomial errors.}
    \label{AgePDFNG}
 \end{figure}

We find two age bins in which there is a (statistical) difference between the CDM and ETHOS gas-poor galaxies, namely  at ages of 600 and 650~Myr. Translated to $z=0$ galaxies, we predict that, for CDM, $>55$~per~cent of gas-starved Local Group dwarfs will have $>10^{5}\msun$ in stellar mass older than 13.3~Gyr as compared to $<40$~per~cent if the dark matter has a power spectrum cutoff at dwarf galaxy scales (roughly at the scale of the benchmark ETHOS model studied here). In practice, testing this prediction is very challenging due to the difficulty in determining the ages of old stellar populations with sub-Gyr precision \citep[see e.g.][]{Weisz14a}.  

Finally, we present a corollary of the increased delay of the formation of ETHOS galaxies (relative to CDM) in progressively lower mass haloes. At a given mass, haloes that form earlier are embedded in larger overdensities (are more clustered) than those that form later \citep{Gao2005}. For CDM, since haloes start collapsing and forming stars very early, 
any difference in the onset of star formation in overdense and underdense regions (at $z=6$) will likely be negligible. 
On the other hand, ETHOS haloes collapse later than their CDM counterparts, and we could therefore expect an enhancement of this effect.

We test this hypothesis by computing the median formation time of the first star particle in each halo, $t_\rmn{fs}$, as a function of the local galaxy number density. We define this density simply as the total number of galaxies -- both centrals and satellites -- located within 1~Mpc (physical) of each halo ($n(<~1{\rm Mpc})$). We calculate this relation for our sample of matched CDM and ETHOS haloes that have halo masses in the range $10^{8}<M_{200}<10^{9}\msun$ and
that have at least one star particle.

\begin{figure}
 	\includegraphics[scale=0.34]{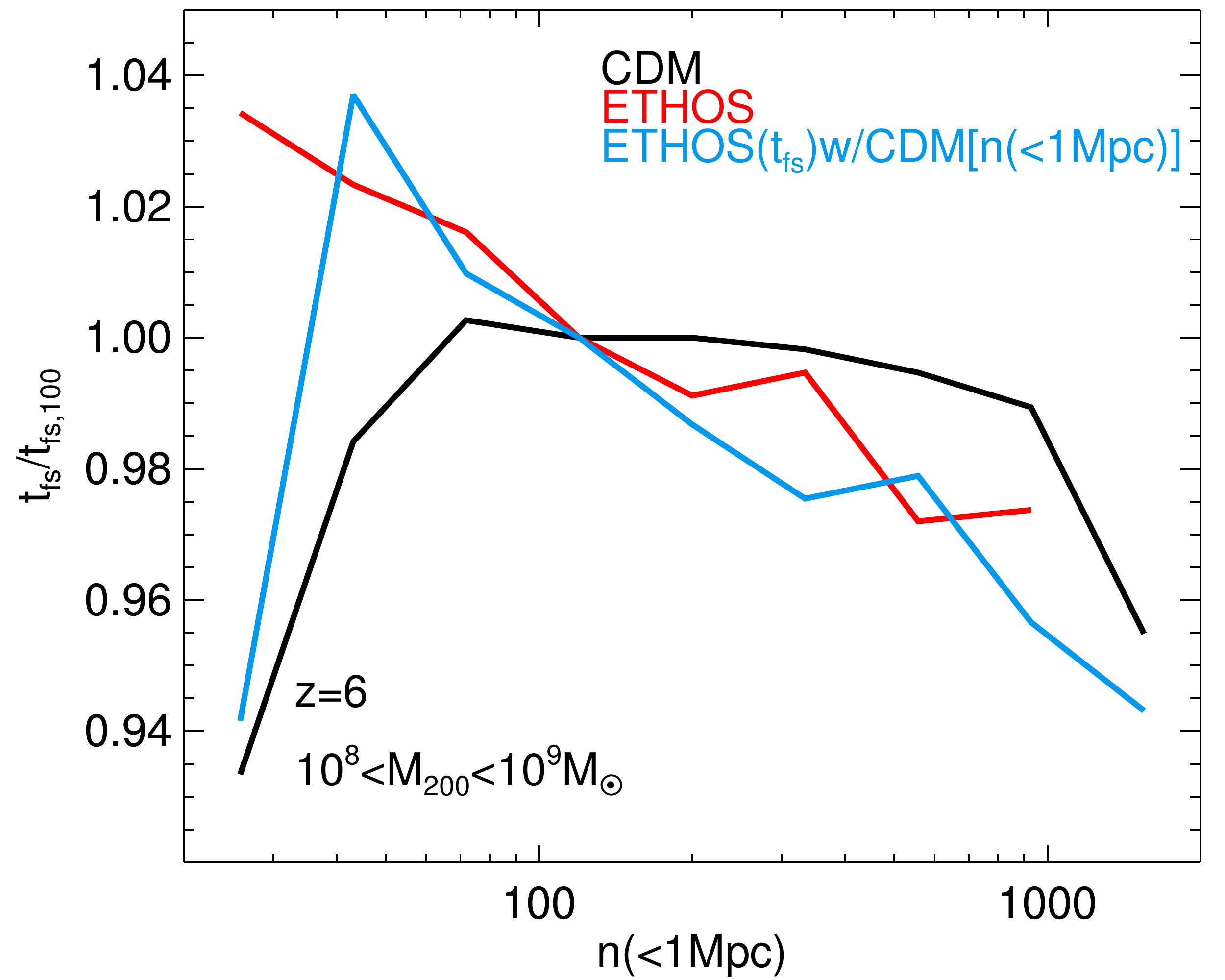}\\
    \caption{Median, normalised time at which the first star particle forms as a function of local galaxy number density (for $z=6$ galaxies). Local density is defined as the number of luminous haloes and subhaloes within a sphere of radius 1~Mpc (physical) centred on a given halo. Each curve is normalised by its value at $n(<1\rmn{Mpc})=100$. The black and red lines are for haloes in the matched CDM and ETHOS samples, respectively. The blue line shows the results of the hybrid calculation in which the ETHOS star formation time is plotted as a function of the CDM local density.}
    \label{TofsN1Mpc}
 \end{figure}

We also include a hybrid calculation in which we take the time $t_\rmn{fs}$ of the ETHOS galaxies with the galaxy number density of the matched CDM counterparts. We normalise $t_\rmn{fs}$ to the value at $n(<~1{\rm Mpc})=100$ in order to focus on the median relation slope, and present our results in Fig.~\ref{TofsN1Mpc}.
 
 We find that the median $t_\rmn{fs}$ -- galaxy number density in CDM (black line) varies by less than two percentage points in CDM between local densities $n(<1~\rmn{Mpc})=[30,1000]$ at $z=6$. At lower densities, the CDM relation turns over which is likely due to small number statistics (20 objects in the lowest bin). Over the same range of galaxy number densities, the ETHOS model results in a much steeper relation (red line), with a median $t_\rmn{fs}$ at $n(<1~\rmn{Mpc})=30$ which is 4~percentage points higher than at $n(<1~\rmn{Mpc})=1000$. The comparison between models is complicated given the overall lower number of galaxies in ETHOS; however, our hybrid calculation (blue line) enables us to `rescale' the $x$-axis for ETHOS haloes. This hybrid relation results in a steeper gradient, which more or less tracks the ETHOS model results, 
with galaxies in the most underdense environments probed forming stars roughly
8~per~cent later than in the densest environments measured at $z=6$. We therefore predict that, in models in which dark matter has a cutoff at the scale of dwarf galaxies, there should be a gradient in the onset of star formation between galaxies residing in high density and low density regions. If this gradient is sufficiently steep, we speculate that ETHOS galaxies in voids form so late that a higher proportion of haloes lose their gas to reionization feedback prior to collapse and are therefore dark, with consequences for the emptiness of the Local Void \citep{Tikhonov09}. 
 
 \section{Conclusions}
 \label{conc}
 
 The nature of dark matter has a vital role to play in the formation of the first galaxies, and this regime provides a complementary probe of new dark matter physics to Local Group observables. In \citet{Lovell18a} we showed that a dark matter model with self-interactions and with a linear matter power spectrum cutoff tuned to alleviate the CDM dwarf-scale challenges in the Local Group 
 -- the ETHOS-4 interacting dark matter model \citep[][hereafter simply `ETHOS']{Vogelsberger16} --  produced a population of galaxies at $z\geq6$, which is consistent with observed high redshift galaxy counts and reionization constraints. This particular ETHOS model has a cutoff very similar to that of a thermal 3.4~keV WDM model, but with the addition of dark acoustic oscillations driven by dark matter-dark radiation interactions in the early Universe. In this follow-up paper we have compared these same simulated ETHOS and CDM galaxies on an individual basis. We are therefore able to describe how halo collapse differs between the two models, and the subsequent effect of this collapse on the baryonic component of the galaxies. 
 
 Our simulations are the same as those of \citet{Lovell18a}, which feature a leading model of galaxy formation \citep[][]{Vogelsberger13, Torrey14, Vogelsberger14,Vogelsberger14b,Genel14, Weinberger17, Pillepich17}. The simulations have a 1024$^3$ particle cosmological box of side length $36$~Mpc with CDM and ETHOS initial conditions. Since the same initial conditions phases were used in CDM and in ETHOS, we were able to use the Lagrangian region matching method of \citet{Lovell14,Lovell18b} to determine to what degree there is a one-to-one correspondence between CDM and ETHOS haloes as a function of halo mass; we then analysed how halo/galaxy properties including halo mass, stellar mass, star formation rate and star formation history change.  
 
We began our analysis by considering the rate at which simulated haloes had bijective matches, defined as our algorithm having matched a given CDM halo to a counterpart in ETHOS and that same ETHOS counterpart being matched back to that CDM halo. We showed that the fraction of CDM haloes with a bijective match in ETHOS is close to unity above a halo virial mass $M_{200}>10M_\rmn{hm}$, where $M_\rmn{hm}$ is the half mode mass in ETHOS (Fig.~\ref{MatchRate}). Around $M_\rmn{hm}$ itself the matched fraction drops precipitously, from 80~per~cent to 40~per~cent for high quality matches, thus indicating that at this mass scale some proportion of haloes are erased by the early Universe dark matter-dark radiation interactions in ETHOS. A second result of this matching analysis was that the ratio between the CDM and ETHOS halo mass functions is very different to the halo match fraction, with the abundance of ETHOS haloes at the $\sim2M_\rmn{hm}$ scale suppressed by 20~per~cent while the bijective matched fraction for CDM is only 5~per~cent below unity. Conversely, the proportion of high quality matches below $M_\rmn{hm}$ was suppressed more strongly than the mass function. We traced this phenomenon back to the fact that the virial masses of low mass ($M_{200}<10\times M_\rmn{hm}$) ETHOS haloes are suppressed relative to CDM: by 50~per~cent at $M_{200}=M_{hm}$ to 80~per~cent at still lower halo masses (Fig.~\ref{M200Match}). 

We repeated variations of this analysis with the stellar mass and found that, remarkably, the switch from CDM to ETHOS changes the stellar mass in a given halo by a median of less than 10~per~cent (Fig.~\ref{SmHm8}). Since the virial mass is suppressed significantly in low-mass ETHOS haloes, this implies that the stellar mass fraction of ETHOS haloes is enhanced by over 10~per~cent at fixed $M_{200}$. Similarly, we found that the gas mass of ETHOS haloes, which sets a fundamental limit on the stellar mass, is the same in ETHOS and CDM at $z=6$ even though the total bound dark matter mass is suppressed by 40~per~cent (Fig.~\ref{MGaD}). The gas mass is even enhanced by 50~per~cent at $M_\rmn{hm}$ in ETHOS relative to CDM at $z=10$, which leads to a boost in the star formation rate at the same redshift of a factor of 2 on a halo-by-halo basis and a factor of 2.5 at fixed halo mass (Fig.~\ref{SFRMatch}, see also \citealp{Bose16c,Lovell18a}.) We demonstrated that the difference between the gas and dark matter distributions was due in part to the extra filamentary material attached to the ETHOS haloes (Fig.~\ref{Expls}) and speculated that much of this extra dark matter would later become unbound.

We explored possible causes for these differences -- or otherwise -- between CDM and ETHOS in the manner of their collapse. It is well known that the power spectrum cutoff delays the time at which perturbations begin to collapse to form haloes \citep{Lovell12,Ludlow16}. We chose to parametrize the onset of halo collapse as relevant for the onset of star formation by the atomic hydrogen `condensation' (cooling) time, which is the time/redshift at which each halo reaches the mass required for atomic hydrogen cooling (approximately $M_{200}=10^{8}\msun$, with a small dependence on redshift.) 

We showed explicitly that the condensation of ETHOS haloes is delayed relative to CDM in a manner that depends on halo mass: ETHOS haloes whose CDM counterparts have $M_{200}>10^{10}\msun$ at $z=6$ condense 50~Myr later than in CDM, whereas those with CDM counterparts of $M_{200}<10^{9}\msun$ were delayed by a median 300~Myr (Fig.~\ref{TimeDiffMatch}). One of the consequences of this behaviour is that, whereas almost all (galaxy-forming) CDM haloes condense in a brief period before $z=10$, and largely even before $z=15$, ETHOS halo condensation time is much more segregated by halo mass, with lower mass haloes delayed progressively longer, relative to CDM. We showed that this change has remarkably little effect on the galaxy stellar mass, and shows a progressively larger suppression in $M_{200}$ with increasing delay time (Fig.~\ref{TimeDiffMatch1}); at the same time we also showed that the luminous fraction of haloes is unaffected by condensation time up to $z=10$, and at later times the luminous fraction is suppressed relative to the CDM value, presumably due to reionization feedback.

The condensation times were found to have some effect on the gas fractions of galaxies at condensation, with ETHOS haloes that condensed at over 400~Myr after the big bang ($z<12$) exhibiting gas fractions 2~per~cent (absolute) lower than other haloes. At earlier condensation times, ETHOS galaxies have gas fractions at condensation that are a median of 2~per~cent higher than CDM,  and there is a significantly higher proportion of ETHOS haloes that have bound gas fractions higher than the universal baryon fraction, possibly due to the way that monolithic collapse proceeds. We also showed that the star formation in $z=6$ ETHOS galaxies is more rapid than their CDM counterparts, forming the first $\sim$50~per~cent of their stars in 23~per~cent less time than in CDM for stellar mass $M_{*}<10^{7}\msun$, and in some cases in under 100~Myr, which is very rare in CDM.

Finally we made predictions for the population of old stars in $z=6$, in practice an integration of the different formation times and prior star formation in CDM and ETHOS. We predicted that in CDM about 90~per~cent of relatively massive ($M_{*}=10^{8}-10^{8.5}\msun$) $z=6$ galaxies should have stellar populations (with more than $10^{5}\msun$ in stars) older than $680$~Myr, compared to 65~per~cent of ETHOS galaxies (Fig.~\ref{AgePDF}). We also considered the case of specifically gas poor, low stellar mass galaxies that could be analogous to fossil Local Group galaxies, and found a statistically significant difference between the two for the proportion of galaxies with stellar populations older than 600~Myr (70~per~cent for CDM versus 30~per~cent for ETHOS, Fig.~\ref{AgePDFNG}. We also showed that, contrary to CDM, the delay in the onset of star formation in low-mass ETHOS galaxies is stronger in underdense regions (measured at $z=6$), and therefore the appearance of a clear age gradient between low mass galaxies in high density and low density regions could provide evidence in favour of a cutoff in the matter power spectrum. 

We have therefore shown the degree to which galaxies form later, and progressively so with decreasing halo mass, in the presence of a cutoff in the linear matter power spectrum. This results in a lower virial mass for the halo, as well as a lower bound dark matter mass. The gas mass is instead enhanced due to a lack of early star formation, leading to a rapid buildup of stellar mass at some later time.

It may therefore be possible to rule out either CDM or ETHOS using a combination of high redshift \citep{Hashimoto18} and Local Group \citep{Weisz14a} dwarf star formation histories, although the former requires excellent spectra of a currently unavailable number of $z>6$ galaxies and the latter requires unpredicted time resolution for the ages of $>12$~Gyr-old stellar populations. Assuming these obstacles can be overcome, the detection of large stellar populations that formed at $z>17$ will rule out ETHOS and dark matter models that feature a similar cutoff, since the halo merger history is incapable of generating haloes of sufficient mass by that time independent of the galaxy formation model. On the other hand, the non-detection of such old populations would heavily favour the presence of a cutoff at the $k\sim20~h\rmn{Mpc}^{-1}$ scale, as without the presence of very early heating from a stellar population it should not be possible to prevent star formation at $z=17$ in CDM.

\section*{Acknowledgements}

We thank Volker Springel for giving us access to {\sc Arepo}. We would like to thank Christoph Pfrommer and Francis-Yan Cyr-Racine for useful comments. MRL is  supported  by  a  COFUND/Durham  Junior Research Fellowship under EU grant 609412. MRL and JZ acknowledge
support by a Grant of Excellence from the Icelandic Research Fund (grant number
173929$-$051). MV acknowledges support through an MIT RSC award, a Kavli Research Investment Fund, NASA ATP grant NNX17AG29G, and NSF grants AST-1814053 and AST-1814259. Some numerical calculations were run on using allocation TG-AST140080 granted by the Extreme Science and Engineering Discovery Environment (XSEDE), which is supported by the NSF.

\bibliographystyle{mnras}

\begin{thebibliography}{}
\makeatletter
\relax
\def\mn@urlcharsother{\let\do\@makeother \do\$\do\&\do\#\do\^\do\_\do\%\do\~}
\def\mn@doi{\begingroup\mn@urlcharsother \@ifnextchar [ {\mn@doi@}
  {\mn@doi@[]}}
\def\mn@doi@[#1]#2{\def\@tempa{#1}\ifx\@tempa\@empty \href
  {http://dx.doi.org/#2} {doi:#2}\else \href {http://dx.doi.org/#2} {#1}\fi
  \endgroup}
\def\mn@eprint#1#2{\mn@eprint@#1:#2::\@nil}
\def\mn@eprint@arXiv#1{\href {http://arxiv.org/abs/#1} {{\tt arXiv:#1}}}
\def\mn@eprint@dblp#1{\href {http://dblp.uni-trier.de/rec/bibtex/#1.xml}
  {dblp:#1}}
\def\mn@eprint@#1:#2:#3:#4\@nil{\def\@tempa {#1}\def\@tempb {#2}\def\@tempc
  {#3}\ifx \@tempc \@empty \let \@tempc \@tempb \let \@tempb \@tempa \fi \ifx
  \@tempb \@empty \def\@tempb {arXiv}\fi \@ifundefined
  {mn@eprint@\@tempb}{\@tempb:\@tempc}{\expandafter \expandafter \csname
  mn@eprint@\@tempb\endcsname \expandafter{\@tempc}}}

\bibitem[\protect\citeauthoryear{{Angulo}, {Hahn}  \& {Abel}}{{Angulo}
  et~al.}{2013}]{Angulo13}
{Angulo} R.~E.,  {Hahn} O.,   {Abel} T.,  2013, \mn@doi [\mnras]
  {10.1093/mnras/stt1246}, \href
  {http://adsabs.harvard.edu/abs/2013MNRAS.434.3337A} {434, 3337}

\bibitem[\protect\citeauthoryear{{Bode}, {Ostriker}  \& {Turok}}{{Bode}
  et~al.}{2001}]{Bode01}
{Bode} P.,  {Ostriker} J.~P.,   {Turok} N.,  2001, \mn@doi [\apj]
  {10.1086/321541}, \href {http://adsabs.harvard.edu/abs/2001ApJ...556...93B}
  {556, 93}

\bibitem[\protect\citeauthoryear{{B{\oe}hm}, {Riazuelo}, {Hansen}  \&
  {Schaeffer}}{{B{\oe}hm} et~al.}{2002}]{Boehm02}
{B{\oe}hm} C.,  {Riazuelo} A.,  {Hansen} S.~H.,   {Schaeffer} R.,  2002,
  \mn@doi [\prd] {10.1103/PhysRevD.66.083505}, \href
  {http://adsabs.harvard.edu/abs/2002PhRvD..66h3505B} {66, 083505}

\bibitem[\protect\citeauthoryear{{B{\oe}hm}, {Schewtschenko}, {Wilkinson},
  {Baugh}  \& {Pascoli}}{{B{\oe}hm} et~al.}{2014}]{Boehm14}
{B{\oe}hm} C.,  {Schewtschenko} J.~A.,  {Wilkinson} R.~J.,  {Baugh} C.~M.,
  {Pascoli} S.,  2014, \mn@doi [\mnras] {10.1093/mnrasl/slu115}, \href
  {http://adsabs.harvard.edu/abs/2014MNRAS.445L..31B} {445, L31}

\bibitem[\protect\citeauthoryear{{Bose}, {Hellwing}, {Frenk}, {Jenkins},
  {Lovell}, {Helly}  \& {Li}}{{Bose} et~al.}{2016a}]{Bose16a}
{Bose} S.,  {Hellwing} W.~A.,  {Frenk} C.~S.,  {Jenkins} A.,  {Lovell} M.~R.,
  {Helly} J.~C.,   {Li} B.,  2016a, \mn@doi [\mnras] {10.1093/mnras/stv2294},
  \href {http://adsabs.harvard.edu/abs/2016MNRAS.455..318B} {455, 318}

\bibitem[\protect\citeauthoryear{{Bose}, {Frenk}, {Hou}, {Lacey}  \&
  {Lovell}}{{Bose} et~al.}{2016b}]{Bose16c}
{Bose} S.,  {Frenk} C.~S.,  {Hou} J.,  {Lacey} C.~G.,   {Lovell} M.~R.,  2016b,
  \mn@doi [\mnras] {10.1093/mnras/stw2288}, \href
  {http://adsabs.harvard.edu/abs/2016MNRAS.463.3848B} {463, 3848}

\bibitem[\protect\citeauthoryear{{Bose} et~al.,}{{Bose} et~al.}{2017}]{Bose17a}
{Bose} S.,  et~al., 2017, \mn@doi [\mnras] {10.1093/mnras/stw2686}, \href
  {http://adsabs.harvard.edu/abs/2017MNRAS.464.4520B} {464, 4520}

\bibitem[\protect\citeauthoryear{{Bose}, {Vogelsberger}, {Zavala}, {Pfrommer},
  {Cyr-Racine}, {Bohr}  \& {Bringmann}}{{Bose} et~al.}{2018a}]{Bose2018}
{Bose} S.,  {Vogelsberger} M.,  {Zavala} J.,  {Pfrommer} C.,  {Cyr-Racine}
  F.-Y.,  {Bohr} S.,   {Bringmann} T.,  2018a, preprint, \href
  {http://adsabs.harvard.edu/abs/2018arXiv181110630B} {} (\mn@eprint {arXiv}
  {1811.10630})

\bibitem[\protect\citeauthoryear{{Bose}, {Deason}  \& {Frenk}}{{Bose}
  et~al.}{2018b}]{Bose18}
{Bose} S.,  {Deason} A.~J.,   {Frenk} C.~S.,  2018b, \mn@doi [\apj]
  {10.3847/1538-4357/aacbc4}, \href
  {https://ui.adsabs.harvard.edu/#abs/2018ApJ...863..123B} {863, 123}

\bibitem[\protect\citeauthoryear{{Bouwens}, {Illingworth}, {Oesch}, {Caruana},
  {Holwerda}, {Smit}  \& {Wilkins}}{{Bouwens} et~al.}{2015}]{Bouwens15}
{Bouwens} R.~J.,  {Illingworth} G.~D.,  {Oesch} P.~A.,  {Caruana} J.,
  {Holwerda} B.,  {Smit} R.,   {Wilkins} S.,  2015, \mn@doi [\apj]
  {10.1088/0004-637X/811/2/140}, \href
  {http://adsabs.harvard.edu/abs/2015ApJ...811..140B} {811, 140}

\bibitem[\protect\citeauthoryear{{Boyarsky}, {Ruchayskiy}  \&
  {Shaposhnikov}}{{Boyarsky} et~al.}{2009}]{Boyarsky09a}
{Boyarsky} A.,  {Ruchayskiy} O.,   {Shaposhnikov} M.,  2009, \mn@doi [Annual
  Review of Nuclear and Particle Science] {10.1146/annurev.nucl.010909.083654},
  \href {http://adsabs.harvard.edu/abs/2009ARNPS..59..191B} {59, 191}

\bibitem[\protect\citeauthoryear{{Boyarsky}, {Franse}, {Iakubovskyi}  \&
  {Ruchayskiy}}{{Boyarsky} et~al.}{2014a}]{Boyarsky14c}
{Boyarsky} A.,  {Franse} J.,  {Iakubovskyi} D.,   {Ruchayskiy} O.,  2014a,
  {arXiv:1408.4388 [astro-ph]}, \href
  {http://adsabs.harvard.edu/abs/2014arXiv1408.4388B} {}

\bibitem[\protect\citeauthoryear{{Boyarsky}, {Ruchayskiy}, {Iakubovskyi}  \&
  {Franse}}{{Boyarsky} et~al.}{2014b}]{Boyarsky14a}
{Boyarsky} A.,  {Ruchayskiy} O.,  {Iakubovskyi} D.,   {Franse} J.,  2014b,
  \mn@doi [Physical Review Letters] {10.1103/PhysRevLett.113.251301}, \href
  {http://adsabs.harvard.edu/abs/2014PhRvL.113y1301B} {113, 251301}

\bibitem[\protect\citeauthoryear{{Bozek}, {Boylan-Kolchin}, {Horiuchi},
  {Garrison-Kimmel}, {Abazajian}  \& {Bullock}}{{Bozek} et~al.}{2016}]{Bozek16}
{Bozek} B.,  {Boylan-Kolchin} M.,  {Horiuchi} S.,  {Garrison-Kimmel} S.,
  {Abazajian} K.,   {Bullock} J.~S.,  2016, \mn@doi [\mnras]
  {10.1093/mnras/stw688}, \href
  {http://adsabs.harvard.edu/abs/2016MNRAS.459.1489B} {459, 1489}

\bibitem[\protect\citeauthoryear{{Bozek} et~al.,}{{Bozek}
  et~al.}{2018}]{Bozek18}
{Bozek} B.,  et~al., 2018, preprint, \href
  {https://ui.adsabs.harvard.edu/#abs/2018arXiv180305424B} {p.
  arXiv:1803.05424} (\mn@eprint {arXiv} {1803.05424})

\bibitem[\protect\citeauthoryear{{Brinckmann}, {Zavala}, {Rapetti}, {Hansen}
  \& {Vogelsberger}}{{Brinckmann} et~al.}{2017}]{Brinckmann17}
{Brinckmann} T.,  {Zavala} J.,  {Rapetti} D.,  {Hansen} S.~H.,   {Vogelsberger}
  M.,  2017, preprint, \href
  {http://adsabs.harvard.edu/abs/2017arXiv170500623B} {} (\mn@eprint {arXiv}
  {1705.00623})

\bibitem[\protect\citeauthoryear{{Buckley}, {Zavala}, {Cyr-Racine}, {Sigurdson}
   \& {Vogelsberger}}{{Buckley} et~al.}{2014}]{Buckley14}
{Buckley} M.~R.,  {Zavala} J.,  {Cyr-Racine} F.-Y.,  {Sigurdson} K.,
  {Vogelsberger} M.,  2014, \mn@doi [\prd] {10.1103/PhysRevD.90.043524}, \href
  {http://adsabs.harvard.edu/abs/2014PhRvD..90d3524B} {90, 043524}

\bibitem[\protect\citeauthoryear{{Bulbul}, {Markevitch}, {Foster}, {Smith},
  {Loewenstein}  \& {Randall}}{{Bulbul} et~al.}{2014}]{Bulbul14}
{Bulbul} E.,  {Markevitch} M.,  {Foster} A.,  {Smith} R.~K.,  {Loewenstein} M.,
    {Randall} S.~W.,  2014, \mn@doi [\apj] {10.1088/0004-637X/789/1/13}, \href
  {http://adsabs.harvard.edu/abs/2014ApJ...789...13B} {789, 13}

\bibitem[\protect\citeauthoryear{{Creasey}, {Sameie}, {Sales}, {Yu},
  {Vogelsberger}  \& {Zavala}}{{Creasey} et~al.}{2017}]{Creasey17}
{Creasey} P.,  {Sameie} O.,  {Sales} L.~V.,  {Yu} H.-B.,  {Vogelsberger} M.,
  {Zavala} J.,  2017, \mn@doi [\mnras] {10.1093/mnras/stx522}, \href
  {http://adsabs.harvard.edu/abs/2017MNRAS.468.2283C} {468, 2283}

\bibitem[\protect\citeauthoryear{{Cyr-Racine}, {Sigurdson}, {Zavala},
  {Bringmann}, {Vogelsberger}  \& {Pfrommer}}{{Cyr-Racine}
  et~al.}{2016a}]{francis16}
{Cyr-Racine} F.-Y.,  {Sigurdson} K.,  {Zavala} J.,  {Bringmann} T.,
  {Vogelsberger} M.,   {Pfrommer} C.,  2016a, \mn@doi [\prd]
  {10.1103/PhysRevD.93.123527}, \href
  {http://adsabs.harvard.edu/abs/2016PhRvD..93l3527C} {93, 123527}

\bibitem[\protect\citeauthoryear{{Cyr-Racine}, {Sigurdson}, {Zavala},
  {Bringmann}, {Vogelsberger}  \& {Pfrommer}}{{Cyr-Racine}
  et~al.}{2016b}]{CyrRacine16}
{Cyr-Racine} F.-Y.,  {Sigurdson} K.,  {Zavala} J.,  {Bringmann} T.,
  {Vogelsberger} M.,   {Pfrommer} C.,  2016b, \mn@doi [\prd]
  {10.1103/PhysRevD.93.123527}, \href
  {http://adsabs.harvard.edu/abs/2016PhRvD..93l3527C} {93, 123527}

\bibitem[\protect\citeauthoryear{{Gao} \& {Theuns}}{{Gao} \&
  {Theuns}}{2007}]{Gao07}
{Gao} L.,  {Theuns} T.,  2007, \mn@doi [Science] {10.1126/science.1146676},
  \href {http://adsabs.harvard.edu/abs/2007Sci...317.1527G} {317, 1527}

\bibitem[\protect\citeauthoryear{{Gao}, {Springel}  \& {White}}{{Gao}
  et~al.}{2005}]{Gao2005}
{Gao} L.,  {Springel} V.,   {White} S.~D.~M.,  2005, \mn@doi [\mnras]
  {10.1111/j.1745-3933.2005.00084.x}, \href
  {http://adsabs.harvard.edu/abs/2005MNRAS.363L..66G} {363, L66}

\bibitem[\protect\citeauthoryear{{Garzilli}, {Boyarsky}  \&
  {Ruchayskiy}}{{Garzilli} et~al.}{2015}]{Garzilli15}
{Garzilli} A.,  {Boyarsky} A.,   {Ruchayskiy} O.,  2015, {arXiv:1510.07006
  [astro-ph]}, \href {http://adsabs.harvard.edu/abs/2015arXiv151007006G} {}

\bibitem[\protect\citeauthoryear{{Garzilli}, {Magalich}, {Theuns}, {Frenk},
  {Weniger}, {Ruchayskiy}  \& {Boyarsky}}{{Garzilli} et~al.}{2018}]{Garzilli18}
{Garzilli} A.,  {Magalich} A.,  {Theuns} T.,  {Frenk} C.~S.,  {Weniger} C.,
  {Ruchayskiy} O.,   {Boyarsky} A.,  2018, preprint, \href
  {https://ui.adsabs.harvard.edu/#abs/2018arXiv180906585G} {p.
  arXiv:1809.06585} (\mn@eprint {arXiv} {1809.06585})

\bibitem[\protect\citeauthoryear{{Genel} et~al.,}{{Genel}
  et~al.}{2014}]{Genel14}
{Genel} S.,  et~al., 2014, \mn@doi [\mnras] {10.1093/mnras/stu1654}, \href
  {http://adsabs.harvard.edu/abs/2014MNRAS.445..175G} {445, 175}

\bibitem[\protect\citeauthoryear{{Hashimoto} et~al.,}{{Hashimoto}
  et~al.}{2018}]{Hashimoto18}
{Hashimoto} T.,  et~al., 2018, preprint, \href
  {http://adsabs.harvard.edu/abs/2018arXiv180505966H} {} (\mn@eprint {arXiv}
  {1805.05966})

\bibitem[\protect\citeauthoryear{{Horiuchi}, {Bozek}, {Abazajian},
  {Boylan-Kolchin}, {Bullock}, {Garrison-Kimmel}  \& {Onorbe}}{{Horiuchi}
  et~al.}{2016}]{horiuchi2016}
{Horiuchi} S.,  {Bozek} B.,  {Abazajian} K.~N.,  {Boylan-Kolchin} M.,
  {Bullock} J.~S.,  {Garrison-Kimmel} S.,   {Onorbe} J.,  2016, \mn@doi
  [\mnras] {10.1093/mnras/stv2922}, \href
  {http://adsabs.harvard.edu/abs/2016MNRAS.456.4346H} {456, 4346}

\bibitem[\protect\citeauthoryear{{Ir{\v s}i{\v c}} et~al.,}{{Ir{\v s}i{\v c}}
  et~al.}{2017}]{Irsic17}
{Ir{\v s}i{\v c}} V.,  et~al., 2017, \mn@doi [\prd]
  {10.1103/PhysRevD.96.023522}, \href
  {http://adsabs.harvard.edu/abs/2017PhRvD..96b3522I} {96, 023522}

\bibitem[\protect\citeauthoryear{{Kennedy}, {Frenk}, {Cole}  \&
  {Benson}}{{Kennedy} et~al.}{2014}]{Kennedy14}
{Kennedy} R.,  {Frenk} C.,  {Cole} S.,   {Benson} A.,  2014, \mn@doi [\mnras]
  {10.1093/mnras/stu719}, \href
  {http://adsabs.harvard.edu/abs/2014MNRAS.442.2487K} {442, 2487}

\bibitem[\protect\citeauthoryear{{Lovell} et~al.,}{{Lovell}
  et~al.}{2012}]{Lovell12}
{Lovell} M.~R.,  et~al., 2012, \mn@doi [\mnras]
  {10.1111/j.1365-2966.2011.20200.x}, \href
  {http://adsabs.harvard.edu/abs/2012MNRAS.420.2318L} {420, 2318}

\bibitem[\protect\citeauthoryear{{Lovell}, {Frenk}, {Eke}, {Jenkins}, {Gao}  \&
  {Theuns}}{{Lovell} et~al.}{2014}]{Lovell14}
{Lovell} M.~R.,  {Frenk} C.~S.,  {Eke} V.~R.,  {Jenkins} A.,  {Gao} L.,
  {Theuns} T.,  2014, \mn@doi [\mnras] {10.1093/mnras/stt2431}, \href
  {http://adsabs.harvard.edu/abs/2014MNRAS.439..300L} {439, 300}

\bibitem[\protect\citeauthoryear{{Lovell} et~al.,}{{Lovell}
  et~al.}{2016}]{Lovell16}
{Lovell} M.~R.,  et~al., 2016, \mn@doi [\mnras] {10.1093/mnras/stw1317}, \href
  {http://adsabs.harvard.edu/abs/2016MNRAS.461...60L} {461, 60}

\bibitem[\protect\citeauthoryear{{Lovell}, {Gonzalez-Perez}, {Bose},
  {Boyarsky}, {Cole}, {Frenk}  \& {Ruchayskiy}}{{Lovell}
  et~al.}{2017a}]{Lovell17a}
{Lovell} M.~R.,  {Gonzalez-Perez} V.,  {Bose} S.,  {Boyarsky} A.,  {Cole} S.,
  {Frenk} C.~S.,   {Ruchayskiy} O.,  2017a, \mn@doi [\mnras]
  {10.1093/mnras/stx621}, \href
  {https://ui.adsabs.harvard.edu/#abs/2017MNRAS.468.2836L} {468, 2836}

\bibitem[\protect\citeauthoryear{{Lovell} et~al.,}{{Lovell}
  et~al.}{2017b}]{Lovell17b}
{Lovell} M.~R.,  et~al., 2017b, \mn@doi [\mnras] {10.1093/mnras/stx654}, \href
  {http://adsabs.harvard.edu/abs/2017MNRAS.468.4285L} {468, 4285}

\bibitem[\protect\citeauthoryear{{Lovell} et~al.,}{{Lovell}
  et~al.}{2018a}]{Lovell18a}
{Lovell} M.~R.,  et~al., 2018a, \mn@doi [\mnras] {10.1093/mnras/sty818}, \href
  {https://ui.adsabs.harvard.edu/#abs/2018MNRAS.477.2886L} {477, 2886}

\bibitem[\protect\citeauthoryear{{Lovell} et~al.,}{{Lovell}
  et~al.}{2018b}]{Lovell18b}
{Lovell} M.~R.,  et~al., 2018b, \mn@doi [\mnras] {10.1093/mnras/sty2339}, \href
  {https://ui.adsabs.harvard.edu/#abs/2018MNRAS.481.1950L} {481, 1950}

\bibitem[\protect\citeauthoryear{{Ludlow}, {Bose}, {Angulo}, {Wang},
  {Hellwing}, {Navarro}, {Cole}  \& {Frenk}}{{Ludlow} et~al.}{2016}]{Ludlow16}
{Ludlow} A.~D.,  {Bose} S.,  {Angulo} R.~E.,  {Wang} L.,  {Hellwing} W.~A.,
  {Navarro} J.~F.,  {Cole} S.,   {Frenk} C.~S.,  2016, \mn@doi [\mnras]
  {10.1093/mnras/stw1046}, \href
  {http://adsabs.harvard.edu/abs/2016MNRAS.460.1214L} {460, 1214}

\bibitem[\protect\citeauthoryear{{Menci}, {Grazian}, {Lamastra}, {Calura},
  {Castellano}  \& {Santini}}{{Menci} et~al.}{2018}]{Menci18}
{Menci} N.,  {Grazian} A.,  {Lamastra} A.,  {Calura} F.,  {Castellano} M.,
  {Santini} P.,  2018, \mn@doi [\apj] {10.3847/1538-4357/aaa773}, \href
  {https://ui.adsabs.harvard.edu/#abs/2018ApJ...854....1M} {854, 1}

\bibitem[\protect\citeauthoryear{{Pillepich} et~al.,}{{Pillepich}
  et~al.}{2018}]{Pillepich17}
{Pillepich} A.,  et~al., 2018, \mn@doi [\mnras] {10.1093/mnras/stx2656}, \href
  {http://adsabs.harvard.edu/abs/2018MNRAS.473.4077P} {473, 4077}

\bibitem[\protect\citeauthoryear{{Planck Collaboration} et~al.,}{{Planck
  Collaboration} et~al.}{2014}]{Planck14}
{Planck Collaboration} et~al., 2014, \mn@doi [\aap]
  {10.1051/0004-6361/201321591}, \href
  {http://adsabs.harvard.edu/abs/2014A%26A...571A..16P} {571, A16}

\bibitem[\protect\citeauthoryear{{Planck Collaboration} et~al.,}{{Planck
  Collaboration} et~al.}{2015}]{PlanckCP15}
{Planck Collaboration} et~al., 2015, {arXiv:1502.01589 [astro-ph]}, \href
  {http://adsabs.harvard.edu/abs/2015arXiv150201589P} {}

\bibitem[\protect\citeauthoryear{{Polisensky} \& {Ricotti}}{{Polisensky} \&
  {Ricotti}}{2011}]{Polisensky2011}
{Polisensky} E.,  {Ricotti} M.,  2011, \mn@doi [\prd]
  {10.1103/PhysRevD.83.043506}, \href
  {http://adsabs.harvard.edu/abs/2011PhRvD..83d3506P} {83, 043506}

\bibitem[\protect\citeauthoryear{{Polisensky} \& {Ricotti}}{{Polisensky} \&
  {Ricotti}}{2014}]{Polisensky14}
{Polisensky} E.,  {Ricotti} M.,  2014, \mn@doi [\mnras]
  {10.1093/mnras/stt2105}, \href
  {http://adsabs.harvard.edu/abs/2014MNRAS.437.2922P} {437, 2922}

\bibitem[\protect\citeauthoryear{{Ritondale}, {Vegetti}, {Despali}, {Auger},
  {Koopmans}  \& {McKean}}{{Ritondale} et~al.}{2018}]{Ritondale18}
{Ritondale} E.,  {Vegetti} S.,  {Despali} G.,  {Auger} M.~W.,  {Koopmans}
  L.~V.~E.,   {McKean} J.~P.,  2018, preprint, \href
  {https://ui.adsabs.harvard.edu/#abs/2018arXiv181103627R} {p.
  arXiv:1811.03627} (\mn@eprint {arXiv} {1811.03627})

\bibitem[\protect\citeauthoryear{{Robertson}, {Ellis}, {Furlanetto}  \&
  {Dunlop}}{{Robertson} et~al.}{2015}]{Robertson15}
{Robertson} B.~E.,  {Ellis} R.~S.,  {Furlanetto} S.~R.,   {Dunlop} J.~S.,
  2015, \mn@doi [\apjl] {10.1088/2041-8205/802/2/L19}, \href
  {http://adsabs.harvard.edu/abs/2015ApJ...802L..19R} {802, L19}

\bibitem[\protect\citeauthoryear{{Schneider}, {Trujillo-Gomez}, {Papastergis},
  {Reed}  \& {Lake}}{{Schneider} et~al.}{2017}]{Schneider17}
{Schneider} A.,  {Trujillo-Gomez} S.,  {Papastergis} E.,  {Reed} D.~S.,
  {Lake} G.,  2017, \mn@doi [\mnras] {10.1093/mnras/stx1294}, \href
  {https://ui.adsabs.harvard.edu/#abs/2017MNRAS.470.1542S} {470, 1542}

\bibitem[\protect\citeauthoryear{{Spergel}, {Flauger}  \& {Hlo{\v
  z}ek}}{{Spergel} et~al.}{2015}]{Spergel15}
{Spergel} D.~N.,  {Flauger} R.,   {Hlo{\v z}ek} R.,  2015, \mn@doi [\prd]
  {10.1103/PhysRevD.91.023518}, \href
  {http://adsabs.harvard.edu/abs/2015PhRvD..91b3518S} {91, 023518}

\bibitem[\protect\citeauthoryear{{Springel}}{{Springel}}{2010}]{Springel10}
{Springel} V.,  2010, \mn@doi [\mnras] {10.1111/j.1365-2966.2009.15715.x},
  \href {http://adsabs.harvard.edu/abs/2010MNRAS.401..791S} {401, 791}

\bibitem[\protect\citeauthoryear{{Springel}, {White}, {Tormen}  \&
  {Kauffmann}}{{Springel} et~al.}{2001}]{Springel01a}
{Springel} V.,  {White} S.~D.~M.,  {Tormen} G.,   {Kauffmann} G.,  2001,
  \mn@doi [\mnras] {10.1046/j.1365-8711.2001.04912.x}, \href
  {http://adsabs.harvard.edu/abs/2001MNRAS.328..726S} {328, 726}

\bibitem[\protect\citeauthoryear{{Tikhonov}, {Gottl{\"o}ber}, {Yepes}  \&
  {Hoffman}}{{Tikhonov} et~al.}{2009}]{Tikhonov09}
{Tikhonov} A.~V.,  {Gottl{\"o}ber} S.,  {Yepes} G.,   {Hoffman} Y.,  2009,
  \mn@doi [\mnras] {10.1111/j.1365-2966.2009.15381.x}, \href
  {http://adsabs.harvard.edu/abs/2009MNRAS.399.1611T} {399, 1611}

\bibitem[\protect\citeauthoryear{{Torrey}, {Vogelsberger}, {Genel}, {Sijacki},
  {Springel}  \& {Hernquist}}{{Torrey} et~al.}{2014}]{Torrey14}
{Torrey} P.,  {Vogelsberger} M.,  {Genel} S.,  {Sijacki} D.,  {Springel} V.,
  {Hernquist} L.,  2014, \mn@doi [\mnras] {10.1093/mnras/stt2295}, \href
  {http://adsabs.harvard.edu/abs/2014MNRAS.438.1985T} {438, 1985}

\bibitem[\protect\citeauthoryear{{Vegetti}, {Despali}, {Lovell}  \&
  {Enzi}}{{Vegetti} et~al.}{2018}]{Vegetti18}
{Vegetti} S.,  {Despali} G.,  {Lovell} M.~R.,   {Enzi} W.,  2018, \mn@doi
  [\mnras] {10.1093/mnras/sty2393}, \href
  {https://ui.adsabs.harvard.edu/#abs/2018MNRAS.481.3661V} {481, 3661}

\bibitem[\protect\citeauthoryear{{Viel}, {Becker}, {Bolton}  \&
  {Haehnelt}}{{Viel} et~al.}{2013}]{Viel13}
{Viel} M.,  {Becker} G.~D.,  {Bolton} J.~S.,   {Haehnelt} M.~G.,  2013, \mn@doi
  [\prd] {10.1103/PhysRevD.88.043502}, \href
  {http://adsabs.harvard.edu/abs/2013PhRvD..88d3502V} {88, 043502}

\bibitem[\protect\citeauthoryear{{Vogelsberger} \& {Zavala}}{{Vogelsberger} \&
  {Zavala}}{2013}]{Vogelsberger2013sidm}
{Vogelsberger} M.,  {Zavala} J.,  2013, \mn@doi [\mnras]
  {10.1093/mnras/sts712}, \href
  {http://adsabs.harvard.edu/abs/2013MNRAS.430.1722V} {430, 1722}

\bibitem[\protect\citeauthoryear{{Vogelsberger}, {Zavala}  \&
  {Loeb}}{{Vogelsberger} et~al.}{2012}]{Vogelsberger12}
{Vogelsberger} M.,  {Zavala} J.,   {Loeb} A.,  2012, \mn@doi [\mnras]
  {10.1111/j.1365-2966.2012.21182.x}, \href
  {http://adsabs.harvard.edu/abs/2012MNRAS.tmp.3127V} {p.~3127}

\bibitem[\protect\citeauthoryear{{Vogelsberger}, {Genel}, {Sijacki}, {Torrey},
  {Springel}  \& {Hernquist}}{{Vogelsberger} et~al.}{2013}]{Vogelsberger13}
{Vogelsberger} M.,  {Genel} S.,  {Sijacki} D.,  {Torrey} P.,  {Springel} V.,
  {Hernquist} L.,  2013, \mn@doi [\mnras] {10.1093/mnras/stt1789}, \href
  {http://adsabs.harvard.edu/abs/2013MNRAS.436.3031V} {436, 3031}

\bibitem[\protect\citeauthoryear{{Vogelsberger} et~al.,}{{Vogelsberger}
  et~al.}{2014a}]{Vogelsberger14}
{Vogelsberger} M.,  et~al., 2014a, \mn@doi [\mnras] {10.1093/mnras/stu1536},
  \href {http://adsabs.harvard.edu/abs/2014MNRAS.444.1518V} {444, 1518}

\bibitem[\protect\citeauthoryear{{Vogelsberger} et~al.,}{{Vogelsberger}
  et~al.}{2014b}]{Vogelsberger14b}
{Vogelsberger} M.,  et~al., 2014b, \mn@doi [\nat] {10.1038/nature13316}, \href
  {http://adsabs.harvard.edu/abs/2014Natur.509..177V} {509, 177}

\bibitem[\protect\citeauthoryear{{Vogelsberger}, {Zavala}, {Cyr-Racine},
  {Pfrommer}, {Bringmann}  \& {Sigurdson}}{{Vogelsberger}
  et~al.}{2016}]{Vogelsberger16}
{Vogelsberger} M.,  {Zavala} J.,  {Cyr-Racine} F.-Y.,  {Pfrommer} C.,
  {Bringmann} T.,   {Sigurdson} K.,  2016, \mn@doi [\mnras]
  {10.1093/mnras/stw1076}, \href
  {http://adsabs.harvard.edu/abs/2016MNRAS.460.1399V} {460, 1399}

\bibitem[\protect\citeauthoryear{{Vogelsberger}, {Zavala}, {Schutz}  \&
  {Slatyer}}{{Vogelsberger} et~al.}{2018}]{Vogelsberger2018}
{Vogelsberger} M.,  {Zavala} J.,  {Schutz} K.,   {Slatyer} T.~R.,  2018,
  preprint, \href {http://adsabs.harvard.edu/abs/2018arXiv180503203V} {}
  (\mn@eprint {arXiv} {1805.03203})

\bibitem[\protect\citeauthoryear{{Wang} \& {White}}{{Wang} \&
  {White}}{2007}]{Wang07}
{Wang} J.,  {White} S.~D.~M.,  2007, \mn@doi [\mnras]
  {10.1111/j.1365-2966.2007.12053.x}, \href
  {http://adsabs.harvard.edu/abs/2007MNRAS.380...93W} {380, 93}

\bibitem[\protect\citeauthoryear{{Weinberger} et~al.,}{{Weinberger}
  et~al.}{2017}]{Weinberger17}
{Weinberger} R.,  et~al., 2017, \mn@doi [\mnras] {10.1093/mnras/stw2944}, \href
  {http://adsabs.harvard.edu/abs/2017MNRAS.465.3291W} {465, 3291}

\bibitem[\protect\citeauthoryear{{Weisz}, {Dolphin}, {Skillman}, {Holtzman},
  {Gilbert}, {Dalcanton}  \& {Williams}}{{Weisz} et~al.}{2014}]{Weisz14a}
{Weisz} D.~R.,  {Dolphin} A.~E.,  {Skillman} E.~D.,  {Holtzman} J.,  {Gilbert}
  K.~M.,  {Dalcanton} J.~J.,   {Williams} B.~F.,  2014, \mn@doi [\apj]
  {10.1088/0004-637X/789/2/147}, \href
  {https://ui.adsabs.harvard.edu/#abs/2014ApJ...789..147W} {789, 147}

\makeatother
\end{thebibliography}

\appendix
  \section{Halo concentration in ETHOS}
  \label{sec:conc}
  
\begin{figure}
 	\includegraphics[scale=0.41]{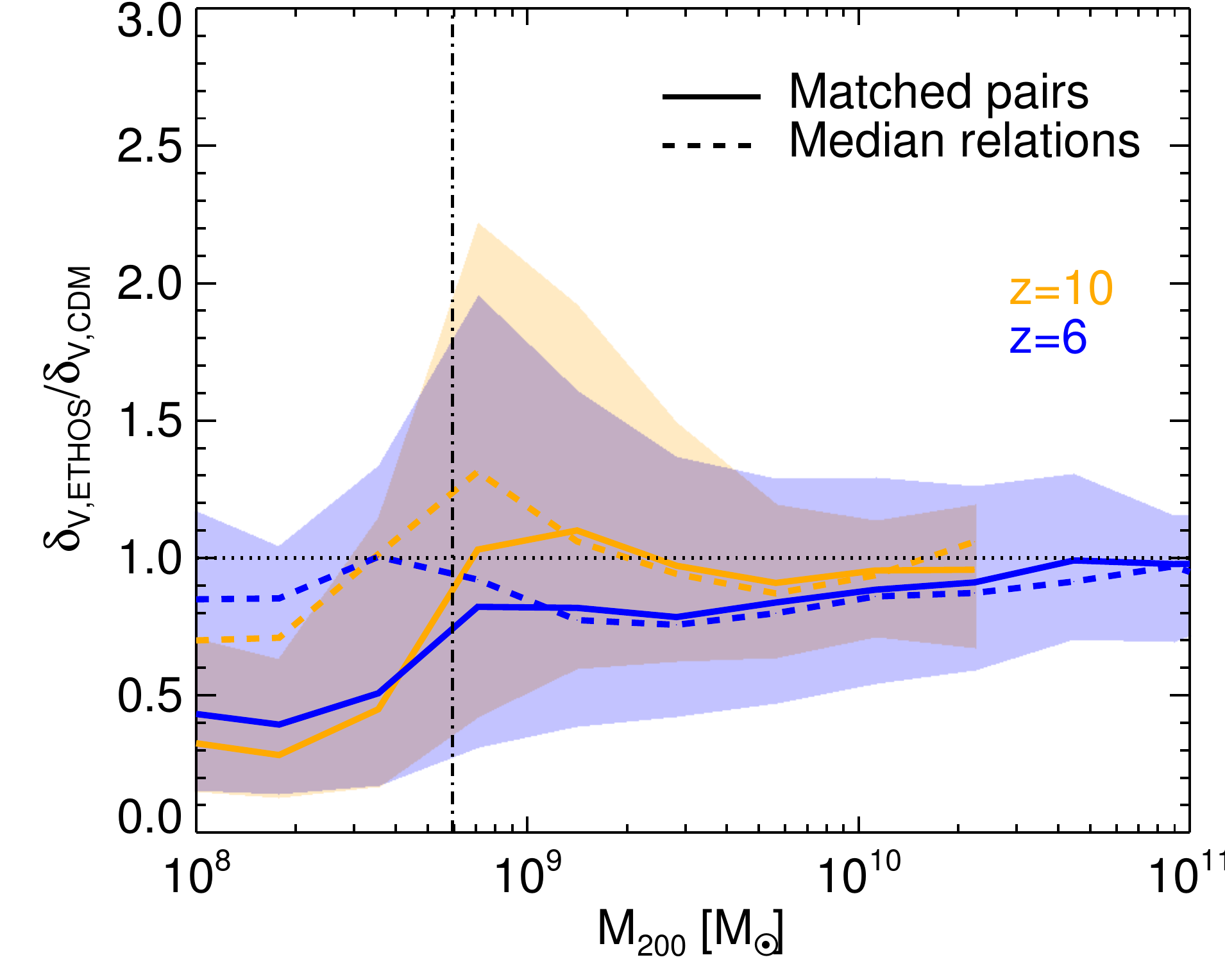}
    \caption{Comparison of $\delta_\rmn{V}$ (a meaure of halo concentration) for matched pairs at $z=6$ (blue) and $z=10$ (orange) in the ETHOS and CDM simulation. The solid lines show the results obtained for the median relation of the ratio of ETHOS and CDM haloes, whereas the dashed line shows the ratio of the two median relations $\delta_\rmn{V}-M_{200}$. The vertical dot dashed line shows the ETHOS half-mode mass.}
    \label{DelVMatch}
 \end{figure}

 The definition of $M_{200}$ is the mass within the radius that encloses an overdensity, $\delta$, $200\times$ the critical density $\rho_\rmn{crit}$, where this radius is denoted $r_{200}$. In order for $M_{200}$ to be lower in ETHOS than in CDM, $r_{200}$ must be smaller and by extension the value of $\delta$ at the halo centre -- as opposed to the $\delta$ measured at $r_{200}$ -- must also be lower. A familiar parametrisation of $\delta$ is given roughly as the ratio of $V_\rmn{max}$ to $r_\rmn{max}$, where the latter quantity is the radius at which $V_\rmn{max}$ occurs. It is defined precisely as:

\begin{equation}
	\delta_{V} = 2\left(\frac{V_\rmn{max}}{H(z)r_\rmn{max}}\right)^2,
\end{equation}
 
 \noindent
where $H(z)$ is the Hubble parameter. This definition of $\delta_{V}$ is also familiar as a profile-independent measure of concentration. We compute the ratio of $\delta_{V}$ for the ETHOS-CDM matched pairs and plot the resulting distributions as a function of $M_{200}$(CDM) in Fig.~\ref{DelVMatch} for $z=6$ and $z=10$. We note however, that since we showed in Fig.~\ref{M200Match} that $M_{200}$ is suppressed in ETHOS for matched pairs, this suppression will contribute towards our results.  We therefore also compute the median $\delta_{V}$ as a function of $M_{200}$ for the ETHOS haloes and separately for the CDM haloes, then compute the ratio of the two medians and plot the results in Fig.~\ref{DelVMatch} as a dashed line. Note that we are still using the haloes that are in the bijectively matched sample for this second calculation, and therefore do not include haloes that are not matched.   
 
 The matched pairs at $z=6$ show two distinct regimes of behaviour in the mass range that we consider well resolved. At the half-mode mass $M_\rmn{hm}$ there is a mild suppression in $\delta_V$ of 20~per~cent from CDM to ETHOS. This suppression gradually disappears towards $M_{200}=10^{11}\msun$, but drops substantially at lower masses. It could be that $M_\rmn{hm}$ is precisely the mass scale at which the mass-concentration relation turns over in ETHOS and leads to lower concentrations at lower masses, although in the COCO simulations this turnover as measured using the NFW concentration, $c$, occurs at $\sim10M_\rmn{hm}$ \citep{Bose16a}. This low-mass dropoff does not occur in the median relation ratio, which is instead relatively flat towards lower masses. This is because the ETHOS halo masses are smaller, and so the haloes here with $M_{200}<10^{9}\msun$ in ETHOS correspond to more massive CDM haloes on the other side of the $10^{9}\msun$ threshold. 
 
 The cutoff below $10^{9}\msun$ also occurs for the $z=10$ haloes. However, the  
haloes more massive than this threshold do not show any suppression in $\delta_{V}$ for ETHOS, unlike the $z=6$ case. The familiar suppression in concentration in models with a power spectrum cutoff is attributed to later formation times of the haloes: this implies that the first objects in ETHOS collapse at the same time as their CDM counterparts, and are those that reside at the site of the highest initial overdensities. This possibility is explored in Section~\ref{sec:ConTimes}.

\begin{figure}
 	\includegraphics[scale=0.41]{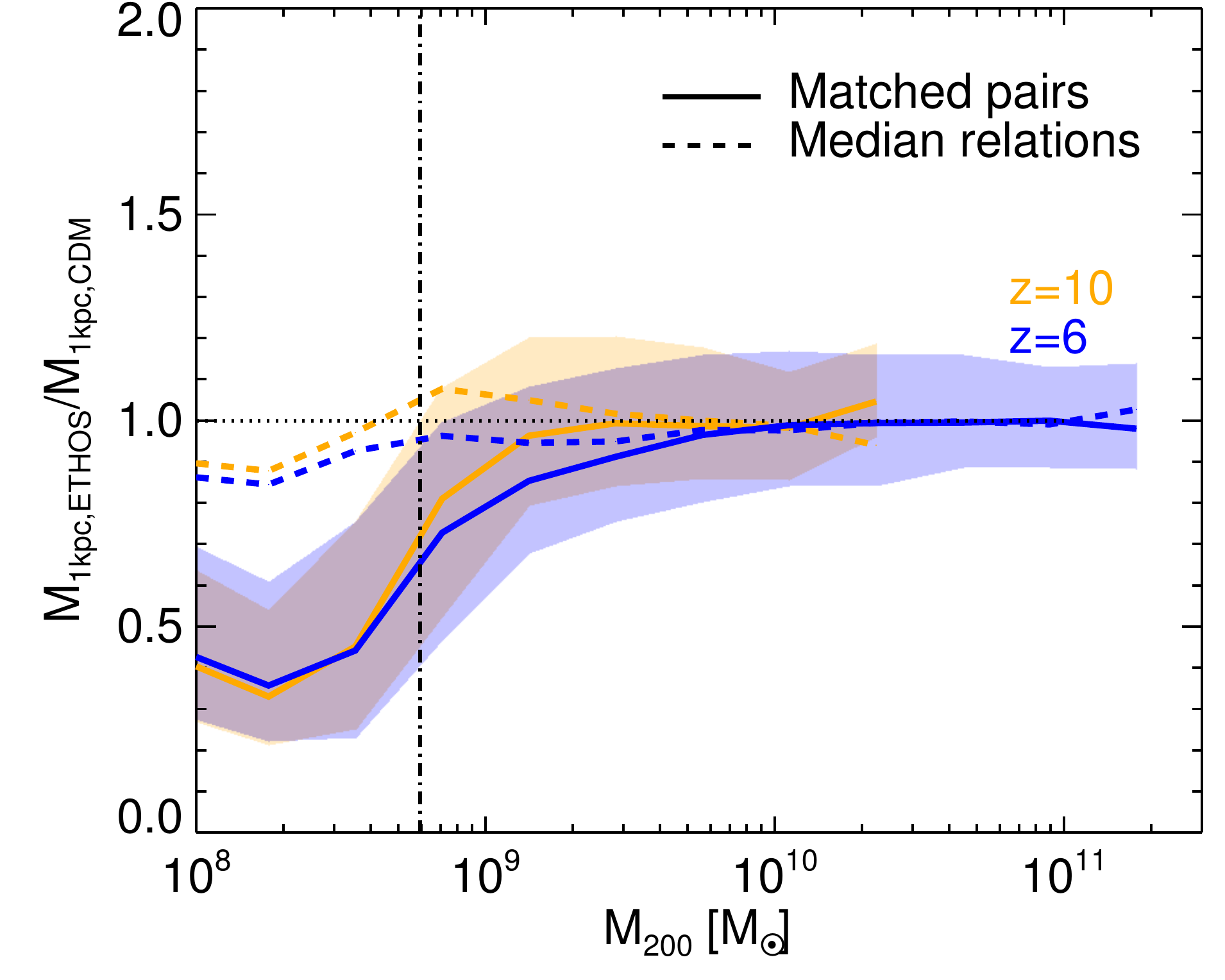}
    \caption{Comparison of dark matter masses within a 1~kpc aperture for matched pairs for $z=10$ (orange) and $z=6$ (blue) in the ETHOS and CDM simulations. Solid lines show median relations of matched pairs and the dashed lines the ratio of the two population medians. The vertical dot-dashed line shows the ETHOS half-mode mass.}
    \label{M3kpcMatch}
 \end{figure}

We end our discussion of the differences in halo formation time with a complementary form of the concentration. The previous measure, $\delta_{V}$, is sensitive to a region of the halo that scales with the virial radius, whereas galaxies typically reside within a smaller region of 1 to 10~kpc, particularly at high redshift. We select a fixed radius of 1~kpc (physical), and compute the ETHOS-CDM ratios as above. The results are presented in Fig.~\ref{M3kpcMatch}.   
 
 The shape of the central dark matter density (defined within 1~kpc) ratio is similar to that using $\delta_{V}$ with a small number of important differences. The suppression of the central density in high-mass ETHOS haloes relative to their CDM counterparts is weaker than $\delta_{V}$ at $z=6$, which shows that the former is dominated by regions further out from the halo centre. For haloes smaller than $3\times10^{9}\msun$ 1~kpc is roughly the size of $r_{200}$ and thus the results are similar. The equality at higher masses could come from two different sources: first, the centres of massive haloes may collapse simultaneously in CDM and ETHOS whereas their outerparts do not, which appears unlikely, or instead there is a degree of mixing within the halo during virialisation that takes place during major mergers that introduces more mass to the halo centre while leaving the outerparts less massive. 
 
 To summarise, we have shown that the progressively lower halo masses in ETHOS are an expression of the lower overdensities of the haloes, and the concentration of ETHOS haloes decreases sharply with respect to CDM for $M_{200}<10^{9}\msun$.
    
\bsp
\label{lastpage}

\end{document}